\documentclass{article}
\usepackage{geometry}
\usepackage{sectsty}
\usepackage{natbib}
\usepackage{hyperref}
\usepackage[all]{hypcap}
\usepackage{graphicx}
\usepackage{placeins}
\usepackage{caption}
\DeclareCaptionLabelFormat{adja-page}{\hrulefill\\#1 #2 \emph{(previous page)}}
\usepackage{wrapfig}
\usepackage{sidecap}
\usepackage{siunitx}
\usepackage{booktabs}
\usepackage{makecell}
\usepackage{amsmath}
\usepackage{xcolor}
\usepackage{xspace}
\hypersetup{colorlinks,
		    linkcolor={blue!40!black},
    		    citecolor={blue!40!black},
		    urlcolor={blue!40!black}}
\newcommand{\beginsupplement}{
        \setcounter{table}{0}
        \renewcommand{\thetable}{S\arabic{table}}
        \setcounter{figure}{0}
        \renewcommand{\thefigure}{S\arabic{figure}}}

\sectionfont{\centering}

\begin{document}

\title{A whitening approach for Transfer Entropy permits the application to narrow-band signals} 

\author{
  Daube, Christoph$^{1*}$\\
    \and
  Gross, Joachim$^{1,2}$\\
    \and
  Ince, Robin A. A.$^1$\\
}

\date{
$^1$Institute of Neuroscience and Psychology, University of Glasgow, 62 Hillhead Street, Glasgow G12 8QB, UK\\
$^2$Institute for Biomagnetism and Biosignalanalysis, University of Münster, Malmedyweg 15, 48149 Münster, Germany\\
*Correspondence: \texttt{christoph.daube@gmail.com, robin.ince@glasgow.ac.uk}
}

\maketitle

\section*{Abstract}
{
\bf
Transfer Entropy, a generalisation of Granger Causality, promises to measure ``information transfer” from a source to a target signal by ignoring self-predictability of a target signal when quantifying the source-target relationship. 
A simple example for signals with such self-predictability are narrowband signals. 
These are both thought to be intrinsically generated by the brain as well as commonly dealt with in analyses of brain signals, where band-pass filters are used to separate responses from noise. 
However, the use of Transfer Entropy is usually discouraged in such cases. 
We simulate simplistic examples where we confirm the failure of classic implementations of Transfer Entropy when applied to narrow-band signals, as made evident by a flawed recovery of effect sizes and interaction delays. 
We propose an alternative approach based on a whitening of the input signals before computing a bivariate measure of directional time-lagged dependency. 
This approach solves the problems found in the simple simulated systems. 
Finally, we explore the behaviour of our measure when applied to delta and theta response components in Magnetoencephalography (MEG) responses to continuous speech. 
The small effects that our measure attributes to a directed interaction from the stimulus to the neuronal responses are stronger in the theta than in the delta band.
This suggests that the delta band reflects a more predictive coupling, while the theta band is stronger involved in bottom-up, reactive processing.
Taken together, we hope to increase the interest in directed perspectives on frequency-specific dependencies.
}

\section*{Introduction}

Over the last decades, the description of statistical dependencies in cerebro-cerebral and cerebro-peripheral pairs of time series has witnessed a surge of interest \citep{bassett2006, brookes2011, naselaris2011, crosse2016, mell2021, gross2021}. 
In these fields, the general idea is to gain insight into the workings of the brain by either studying how time series of neuronal activity relate to other neuronal activity or to external signals such as auditory or visual stimuli as well as the activity of other organs.

As a consequence, countless methodological approaches have been suggested to mathematically quantify these dependencies \citep{bastos2016}. 
Some of these ideas specifically aim at the description of directed interactions, for example by using measures of the so-called “Granger-causal” \citep{granger1969} family, or their generalisation to nonlinear relationships, Transfer Entropy \citep{schreiber2000, barnett2009}.
In both of these measures, the main idea is to quantify a directional dependency by first assessing to what degree a target time-series can be predicted from itself and secondly assessing to what degree this auto-prediction can be improved upon with the assumed source time-series. 
In its classic formulation, TE implements this by way of conditioning the mutual information (MI) between source and target on an operationalisation of the target past. 
It has been suggested that this warrants the capacity to correctly estimate not only ``predictive information transfer” between source and target time-series, but also the recovery of the true underlying interaction delay \citep{wibral2013}. 
The precise estimation of such quantities is of high interest for the research programmes not only of functional connectivity, but of cognitive neuroscience in general.

However, in the arguably simplest and in many applications ubiquitous case of self-predictable or auto-correlated time-series, namely narrowband time-series as obtained e.g. when applying band-pass filters, TE fails to deliver intuitively comprehensible results. 
The application of TE in such cases has therefore repeatedly been discouraged \citep{florin2010, barnett2011}. 

Such applications however are of potentially high interest, given that frequency specific interactions are at the core of popular hypotheses about cerebro-peripheral \citep{giraud2012, donhauser2020} and cerebro-cerebral \citep{schnitzler2005, fries2015, michalareas2016, schoffelen2017} interactions. 
There is an abundance of evidence for intrinsically auto-correlated or band-limited parts of neuronal activity (``oscillations”), whose presence in neuronal recordings \citep{wang2010, donoghue2020} should accordingly impede the use of TE even without the use of analysis filters. 
Moreover, in light of the usually low signal-to-noise ratio (SNR) of many recording modalities, particularly of non-invasive neuroimaging, the isolation of band-limited activity via spectral filtering is a pervasive strategy to achieve acceptable sensitivity and specificity. 
The main idea of TE thus turns out to be an empty promise for many real-world applications where auto-correlations are indeed clearly visible. 

Suggestions to implement TE mainly differ in their approaches to remove the self-predictability of the target signal from the quantification of the effect. 
While it has been argued previously that simplistic approaches relying on a target past operationalisation consisting of a single delay are insufficient \citep{wibral2013}, such approaches remain popular. 
This might imply that the literature is missing more intuitive demonstrations of the shortcomings of such approaches.
Furthermore, it also has been shown previously that prominent proposals relying on multidimensional embeddings of the target time series fail in scenarios of narrow-band effects \citep{wollstadt2017}. 
While suggestions exist that try to overcome this problem by means of constructing frequency-specific surrogate data \citep{pinzuti2020} or state-space models \citep{faes2017}, more intuitive explanations of these failure cases are arguably still lacking. 
Such an intuitive understanding should pave the way towards both more widespread awareness as well as simple fixes of these issues.

One interesting use case for measures that overcome the problems outlined above is the heavily studied phenomenon of speech envelope tracking as observed in magneto- and electroencephalography (henceforth MEEG to denote both modalities) recordings \citep{ahissar2001, hertrich2012, gross2013, ding2012, osullivan2015, diliberto2015, wostmann2017, brodbeck2018b, daube_simple_2019, obleser2019, zan2020, donhauser2020}. 
In short, the low-frequency portion of MEEG signals is reliably related to the time varying energy of the speech signal at a certain delay. 
In noninvasive recordings, speech envelope tracking is heavily studied in the canonical delta and theta bands \citep{ding2014}. 
At higher frequencies, the effect usually fails to robustly exceed noise thresholds \citep[but see][]{kulasingham2020}.
Limiting the analysis to the delta and theta frequency bands is thus an efficient means to achieve stronger effects.
Theories exist that link this phenomenon to frequency specific mass-neuronal processes of debated algorithmic significance \citep{giraud2012, hyafil2015}. 

From the perspective of directed connectivity, an interesting question is to what degree the resemblance of MEEG signals and the speech envelope at a given time $t$ can be explained from the MEEG signal’s own past, and to what degree it can only be explained when additionally considering the stimulus. 
Given the auto-correlated nature of both neuronal processes and the speech envelope in the relevant spectral range, it is in theory possible for a system to predict the upcoming speech envelope \citep[and, as made evident by the progress of past and current machine learning approaches, also richer parts of the speech signal][]{chung2020, lakhotia2021}.
According to popular theories of brain function, such predictive coding is also of high utility for a biological system \citep{rao1999, friston2005}. 
The extent to which speech tracking as quantified by undirected measures such as delayed MI would decrease when using suitable directed measures would highlight the extent to which the heavily studied tracking might in fact reflect predictive rather than reactive processing. 
This would add to accounts that characterise low-frequency oscillations as the deliberate effort of biological systems to be in a state of optimal neuronal excitability \citep{lakatos2008, henry2012}, such that metabolically costly states of high encoding fidelity co-occur with the relevant parts of the stimulus \citep{jones1989, schroeder2009, kayser2015, mlynarski2018b}.

Here, we consider a simple simulated system of band-limited delayed bivariate interactions with a clear and intuitively comprehensible ground truth spectral range and delay. 
We implement delayed MI, two classic algorithms to estimate TE as well as a novel, whitening based estimator (``Directed Information based on conditional entropy”, ``DI\textsubscript{ce}”) within the Gaussian copula MI framework \citep[gcmi,][]{ince2017b} and extensively test these measures with simulations. 
Finally, we explore the behaviour of DI\textsubscript{ce} in an MEG dataset of continuous speech listening.
We find that estimates of the delay between the stimulus and the response as well as the recovered interaction strength in the delta and theta bands differ from those recovered by the bivariate delayed MI.

\section*{Results}

\subsection*{An intuitive overview over multiple measures}

A first goal of this study was to provide an overview of various implementations of TE in a simple and intuitively understandable example case. 
To do so, we simulated coupled systems with a known ground truth spectral interaction range and delay. 
These consisted of 4 - 8 Hz narrow-band filtered Gaussian white noise to obtain a band-limited source signal, which was delayed in time by .15 s to obtain a target signal. 
We added Gaussian white noise to these signals to mimic the noisy measurement process in neuroimaging. 
For such a simplistic but intuitive system, a suitable measure of TE should peak at the ground truth simulated delay and spectral range. 
Further, for such highly auto-correlated signals, TE should yield a highly reduced effect size in comparison to undirected measures such as delayed MI: a potential source signal can only add little information if the target signal is already highly predictable from its own recent history.

\begin{figure}[!ht]
\centering
\includegraphics[width=\linewidth, trim={2.9cm 2.35cm 1.2cm 1.65cm},clip]{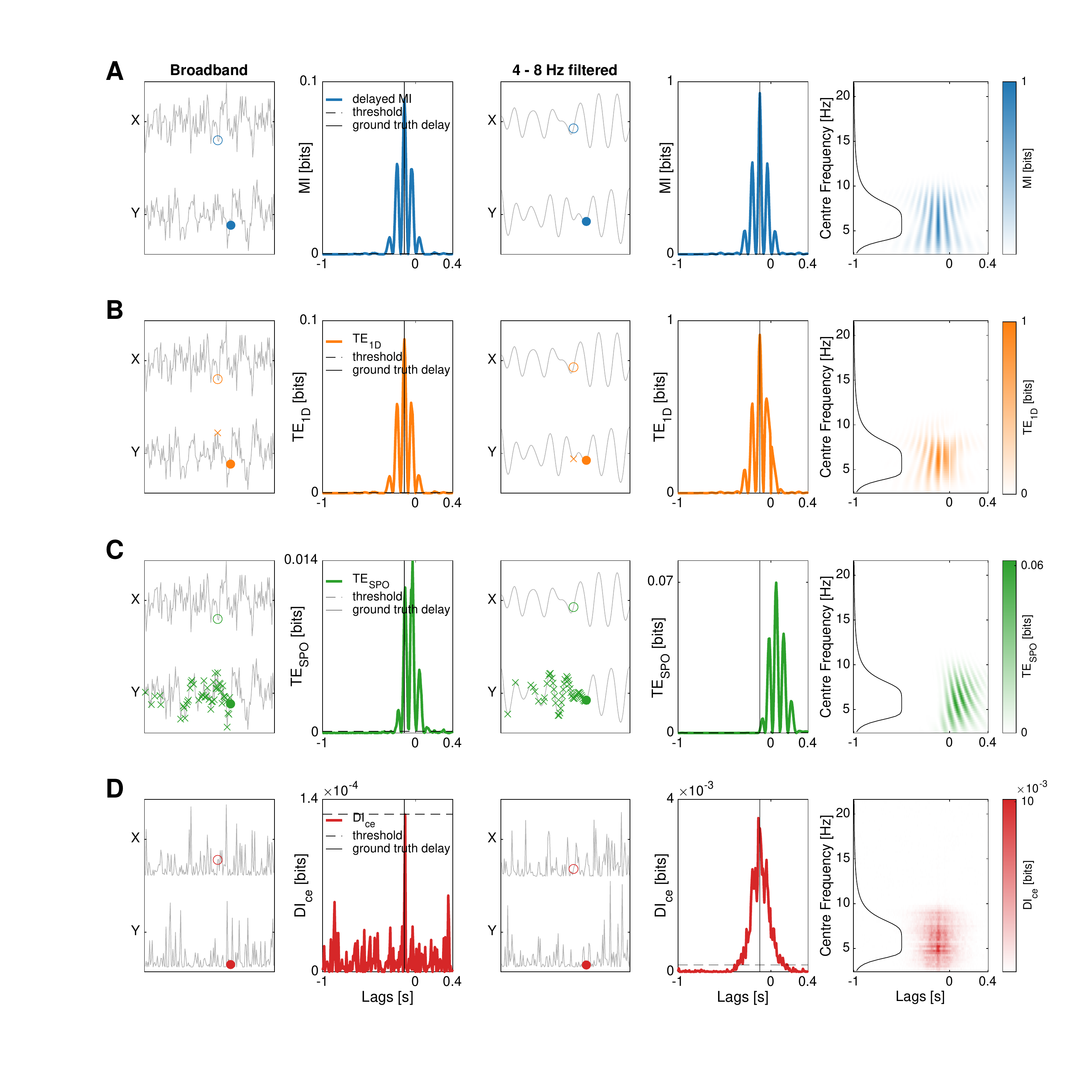}
\caption[Comparison of delay profiles of various undirected and directed dependency measures in a simplistic simulation scenario.]{\textbf{Comparison of delay profiles of various undirected and directed dependency measures in a simplistic simulation scenario.}

The simulation here consists of a 4 - 8 Hz bandpass filtered white noise source signal that is temporally delayed to obtain the target. 
White noise is added to both source and target to model measurement noise. 
First and middle columns show time domain sections of the analysed time series, where a filled circle denotes a target present sample, an empty circle denotes a source past sample at a delay corresponding to the simulated effect, and crosses denote the respective target past samples. 
2nd and 4th columns show delay profiles of the respectives measures relative to the ground truth interaction delay and noise thresholds.
5th column shows spectrotemporal maps, where the frequency response of the filter used to generate the ground truth effect is overlaid as a black curve. 
\textbf{A} Delayed Mutual Information correctly recovers the ground truth interaction delay. 
\textbf{B} TE\textsubscript{1D} also correctly recovers the interaction delay, but measures an interaction effect that is highly similar to delayed MI. 
\textbf{C} TE\textsubscript{SPO} recovers an interaction delay different from the ground truth, even more so when the analysis is performed on the filtered signal. 
\textbf{D} DI\textsubscript{ce} only finds a super-threshold effect with correct recovery of the interaction delay when an analysis filter is used, but with a strongly reduced effect size in comparison to other measures.
}
\label{fig_te_overview}
\end{figure}

\hyperref[fig_te_overview]{Figure \ref{fig_te_overview}} shows the results of applying delayed MI, classical TE estimators TE\textsubscript{1D} and TE\textsubscript{SPO} as well as our proposal, DI\textsubscript{ce} (see below for more detailed explanations of each individual measure) to the same time series, both without and with applying an analysis filter.

A first observation is that in the present implementation with gcmi \citep{ince2017b}, all measures yield increased effect sizes when a suitable analysis filter is applied. 
This, potentially trivially \citep[but see][]{pinzuti2020}, demonstrates that the effect size and consequently the sensitivity can benefit from the application of analysis filters.
It thus underscores the utility of directed connectivity measures that behave robustly when applied to narrow-band signals. 
Further, the pass-band of the effect can be found by applying analysis filters of varying centre frequencies: all measures return the highest effect sizes across analysis bands within the pass-band of the simulated effect.

\graphicspath{{./Figures/}}

Secondly, we observe that TE\textsubscript{1D} fails to return a reduced effect size in comparison to the undirected delayed MI, even when applied to highly self-predictable signals as simulated here. 
This happens for both broadband and filtered analysis scenarios. 
TE\textsubscript{1D} is a classical implementation of TE \citep{besserve2010, lobier2014, ince2015, park2015, giordano2017, morillon2017} that operationalises the target past by means of using only one delay of the target time series (more specifically, the same delay as that of the source signal relative to the target present when scanning across delays).
For the ground truth delay simulated here, this single delay fails to capture most of the self-predictability of the target signal, resulting in an overestimation of the directed effect.

A third measure, TE\textsubscript{SPO} \citep{wibral2013}, yields the anticipated strong reduction in effect size by roughly an order of magnitude in comparison to MI. 
It achieves this by using a more effective handle on the target past: a multi-dimensional “non-uniform” \citep{vlachos2010, faes2011} embedding optimised for self-prediction. 
As opposed to TE\textsubscript{1D}, this essentially consists of multiple delays that are independent of the analysis delay in the scanning procedure. 
However, it fails to recover the ground truth interaction delay simulated in the coupled systems \citep{wollstadt2017}, in both broadband and filtered analysis settings. 
Given this failure in a simplistic problem setting that is ubiquitous in neuroscientific datasets, it is unclear to which degree TE\textsubscript{SPO} can practically live up to its promise of correctly quantifying ``predictive information transfer” as well as the interaction delay.

Finally, the last measure considered in and proposed by this study, DI\textsubscript{ce}, solves both of these problems: It returns the smallest effect sizes in this comparison, even to the degree that it fails to detect the effect in a broadband analysis setting. 
However, when a suitable analysis filter is applied, it returns a delay profile with a super-threshold peak at the simulated delay. 
It achieves this in a two-step approach \citep{haugh1976}:
First, both source and target signals are transformed into time series of surprisal, i.e. the sample-wise entropy of each time point conditional on the same non-uniform embedding used in TE\textsubscript{SPO}. 
Secondly, delayed MI is computed for these whitened time series.

Taken together, this first analysis provides an overview of four different connectivity measures in an intuitive and simplistic simulation setting, illustrating how DI\textsubscript{ce} succeeds in returning intuitive results where TE\textsubscript{1D} and TE\textsubscript{SPO} fail.

\subsection*{Synergy of source and target past about target present distorts conditional MI based TE}

\begin{figure}[!ht]
\centering
\includegraphics[width=\linewidth, trim={3cm 1.2cm 3.05cm 1.5cm},clip]{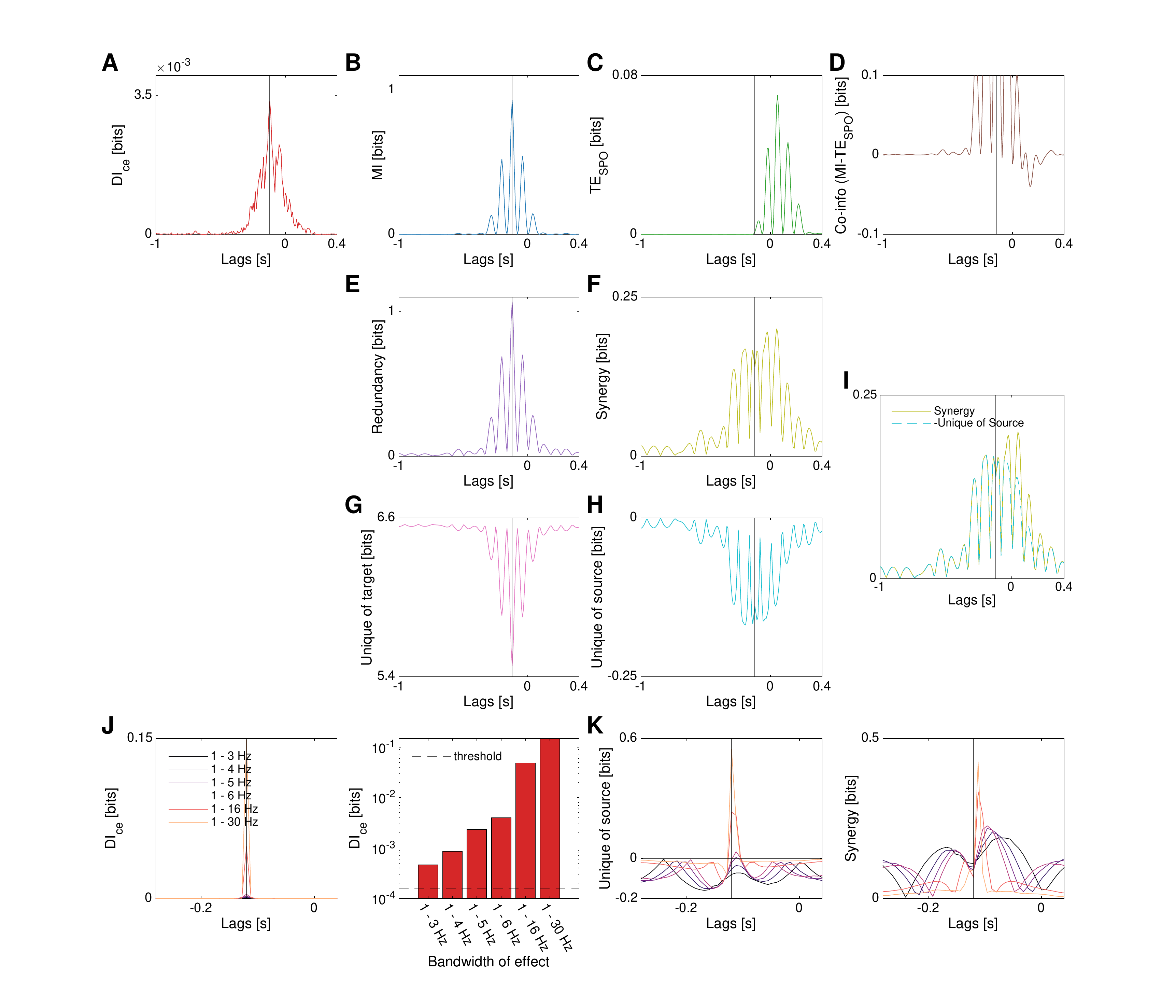}
\caption[Partial information decomposition perspective on TE.]{\textbf{Partial information decomposition perspective on TE.}

\textbf{A} Delay profile of DI\textsubscript{ce} for reference. 
\textbf{B} Delay profile of delayed MI. 
\textbf{C} Delay profile of TE\textsubscript{SPO}, which is based on conditional MI and can, according to PID, be seen as the sum of synergy (F) and unique information of the source (H). 
\textbf{D} Delay profile of co-information (abbreviated as “co-info”), which quantifies the triple set intersection in classic Venn-diagram conceptualisations of three variable systems. 
According to PID, it is the net sum of synergy (F) and redundancy (E). 
Axis limits chosen to highlight the negative (net synergistic) portion at positive delays. 
\textbf{E - H} Delay profiles of the PID atoms redundancy, synergy and unique information (of target and source). 
\textbf{I} Delay profiles of synergy and sign-flipped unique information of the source, highlighting a surplus of synergy at positive delays. 
\textbf{J} Delay profiles of DI\textsubscript{ce} when simulating data with different ground truth effect bandwidths. 
DI\textsubscript{ce} recovers the correct delay at all bandwidths (left) and recovers an increasing effect size as a function of bandwidth. 
It exceeds the noise threshold in all cases. 
\textbf{K} Delay profiles of unique information of the source (left) and synergy (right) for the same effect bandwidths as in J. 
For narrow bandwidths, unique information of the source fails to become positive and recovers the wrong delay.
Further, the sum of unique information of the source and synergy is dominated by synergy, which features a greater delay estimation error than unique information.
}
\label{fig_te_pid}
\end{figure}

We wanted to investigate possible explanations for the counterintuitive results returned by our implementation of TE\textsubscript{SPO}. 
To do so, we considered a perspective on TE offered by partial information decomposition (PID). 
PID is an information theoretic approach to study trivariate relationships \citep{williams2010, ince2017}. 
The central goal in PID is to quantify shared information (redundancy) between two source variables about a third target variable. 
Further, PID aims to measure unique information of both source variables as well as synergistic information that is only obtainable when jointly considering both sources. 
A key insight of PID relevant to TE is that the basis of TE, conditional MI, is the sum of two ``atoms” of PID: unique information and synergy \citep{james2016}.
In other words, conditioning a bivariate relationship on a third variable will remove redundant information, but will not deliver solely unique information. 
It has been pointed out that this conflation of unique and synergistic information defies interpretations of TE as measuring ``information flow” \citep{james2016}. 
By relying on conditional MI, TE measures not just unique information of the source about the target present ignoring the target past, but instead conflates this with synergistic information stemming from interactions of the source and the target past. 
The resulting quantity is thus not ``localisable” to the source \citep{james2016}.

We turned to data from the same simulated system as analysed in the previous section. 
In a first step, we considered co-information \citep{mcgill1954} of the source, the target present and the target past.
It can be obtained by subtracting conditional MI from bivariate MI, and thus quantifies the triple set intersection in classic Venn diagram conceptualisations. 
It can take on positive and negative values, where positive values denote redundancy and negative values denote synergy. 
It is important to note that from a PID perspective, the conceptually simpler (and less controversial) co-information conflates redundancy and synergy to a single net quantity, which PID aims to decompose into pure redundancy and synergy. 
\hyperref[fig_te_pid]{Figure \ref{fig_te_pid}D} shows negative co-information values at later delays, demonstrating that there is indeed a net synergy of the source and the target past about the target present.

Next, we applied PID using an implementation based on common change in surprisal \citep[I$_{ccs}$][]{ince2017}. 
We find that TE\textsubscript{SPO} (\hyperref[fig_te_pid]{figure \ref{fig_te_pid}C}), as it is based on conditional MI, can be decomposed into largely identical profiles of synergy (\hyperref[fig_te_pid]{figure \ref{fig_te_pid}F}) and unique information of the source (\hyperref[fig_te_pid]{figure \ref{fig_te_pid}H}) differing only in their sign. 
The positive net conditional MI (i.e. TE\textsubscript{SPO}) however turns out to entirely stem from a surplus of synergy at later delays that is not cancelled out by unique information of the source (\hyperref[fig_te_pid]{figure \ref{fig_te_pid}I}). 
In other words, in this simplistic example case, TE\textsubscript{SPO} measures a synergistic effect.

As a consequence of the PID perspective on conditional MI based TE implementations, it has been proposed to use the unique information of the source as a more appropriate measure in order to avoid the quantification of synergistic effects \citep{barrett2015}. 
However, in our example case, this quantity is negative across the entire range of considered interaction delays (\hyperref[fig_te_pid]{figure \ref{fig_te_pid}H}), meaning that from considering the source on its own, predictions of the target would become worse. 
This can be seen as the result of two factors: firstly, in this simulation, the source signal is noisy, and secondly, the simulated effect has a very narrow spectral range and thus highly limited degrees of freedom. 
As these factors come together, the efficient operationalisation of the target past makes it impossible to improve on its prediction of the target present. 

We thus considered similar simulations with varying bandwidths of the ground truth effect, and verified that at broader bandwidths, the unique information of the source becomes positive (\hyperref[fig_te_pid]{figure \ref{fig_te_pid}K}). 
Of note, the delay recovered by unique information of the source generally has lower deviations from the ground truth than what is recovered by synergy, but this deviation is still non-zero at narrower effect bandwidths. 
We further find that with increasing bandwidths of the effect, synergy increases as well and peaks at delays closer to the ground truth delay. 
Crucially however, across all simulated bandwidths, our proposed measure, DI\textsubscript{ce}, finds super-threshold effects at the correct delay (\hyperref[fig_te_pid]{figure \ref{fig_te_pid}J}). 

Taken together, we have shown that the delay mis-estimation problems of TE\textsubscript{SPO}, when applied to narrow-band signals, stem from the known conflation of unique and synergistic effects in conditional MI based TE implementations. 
In this example, the effect quantified by TE\textsubscript{SPO} is in fact dominated by synergy. 
We further show that a possible alternative, namely unique information of the source as obtained from PID, has both a lower sensitivity and a worse performance in recovering the ground truth interaction delay than our proposed conditional MI-free measure DI\textsubscript{ce}.

\subsection*{Varying the ground truth interaction delay}

After these initial intuitive and single sample based demonstrations of problems of two classic algorithms to quantify TE as well as a suggestion to explain their origin, we were interested in more thoroughly testing the characteristics of the measures. 
To do so, we performed extensive analyses of the simplistic coupled systems that measured the problems over repeated samples. 
For each of these samples, we computed the delay profiles of the set of four measures and evaluated the recovered interaction delay as well as the recovered effect size as obtained from the peaks across the delay profile of a given repetition. 

\begin{figure}[!ht]
\centering
\includegraphics[width=.75\linewidth, trim={.5cm 1.2cm 1.75cm .9cm},clip]{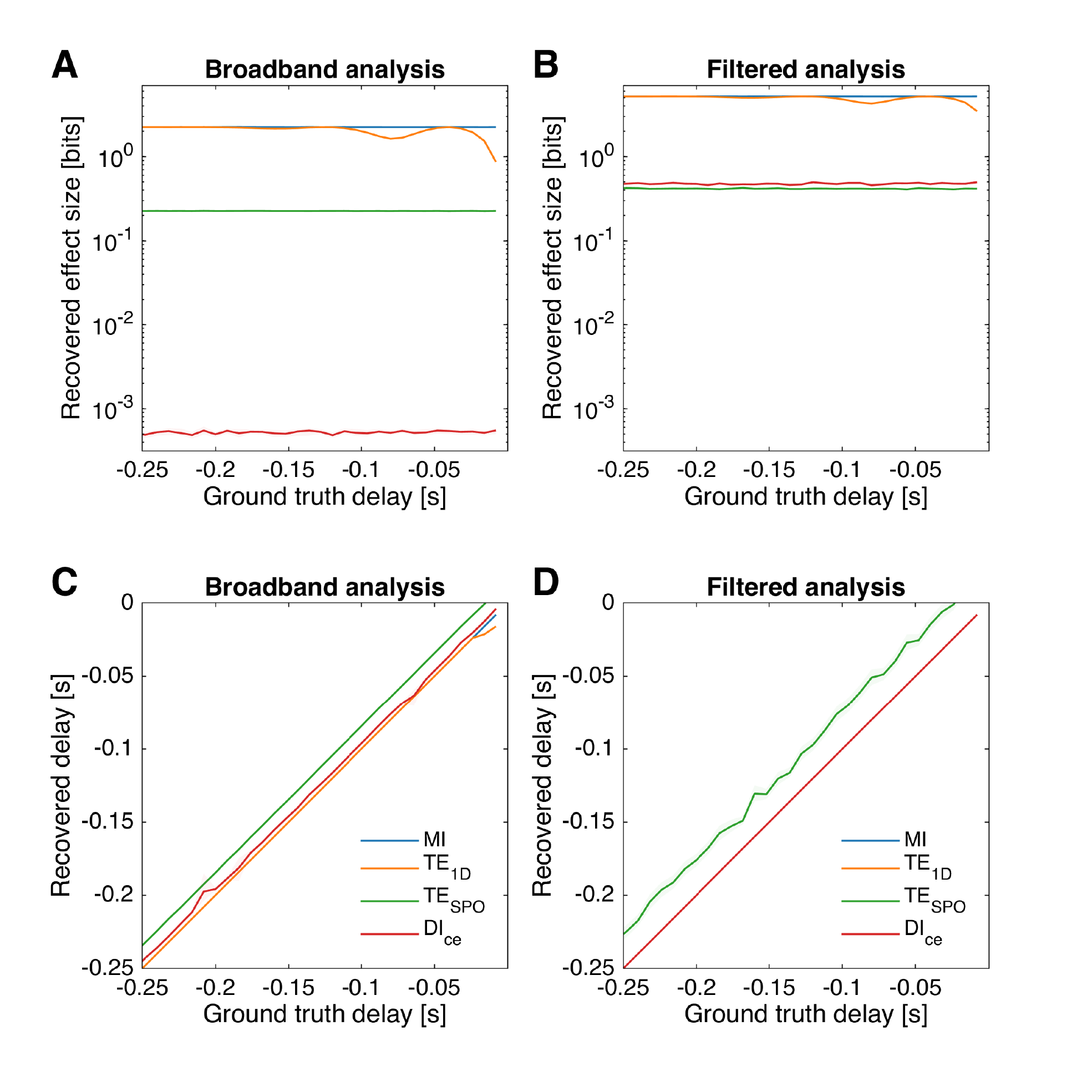}
\caption[Simulation with varying ground truth delay.]{\textbf{Simulation with varying ground truth delay.}

Basic setup of the simulation is the same as in \hyperref[fig_te_overview]{figure \ref{fig_te_overview}}, however, here, the ground truth interaction delay is varied, and 100 repetitions are sampled. 
Moreover, a higher SNR is used. 
Plots show the median across repetitions, shaded regions indicate bootstrapped 95\% confidence intervals. 
\textbf{A} Recovered effect size when the data are analysed without a filter. 
TE\textsubscript{1D} suffers from recovering a systematically varying effect size across different ground truth delays despite a constant simulated interaction strength. 
\textbf{B} Same as A, but applying an analysis filter. 
\textbf{C} Recovered delays when data are analysed without a filter. 
TE\textsubscript{SPO} underestimates the true interaction delay. 
\textbf{D} Same as C, but applying an analysis filter. 
}
\label{fig_te_delay}
\end{figure}

In a first simulation, we were interested in the characteristics of the measures across a range of simulated ground truth interaction delays. 
In principle, a suitable measure of TE should recover the same constant effect size at varying simulated interaction delays when all other factors are kept constant. 
In \hyperref[fig_te_delay]{figure \ref{fig_te_delay}}, the results demonstrate that this is indeed the case for all measures in both broadband and filtered analysis scenarios except for TE\textsubscript{1D}. 
This measure exhibits a systematic variation in the recovered effect size across different simulated interaction delays reminiscent of the filter ringing of the auto-correlation profile of the target signal. 
The failure here again highlights the problems of operationalising the target past with a single delay that varies as the analysis delay is scanned \citep{wibral2013}. 
In case of very short ground truth delays, the MI at the peak of the delay profile is conditioned on the same short delay of the target variable. 
Since narrow-band signals have high auto-correlation at such short delays, this leads to a strong reduction of conditional MI vs MI. 
As the ground truth delay increases, the MI is conditioned on longer delays, where the auto-correlation of the target variable wanes and waxes and thus leaves a correspondingly varying conditional MI. 
This can lead to potential interpretational pitfalls when for example the results of TE\textsubscript{1D} obtained in two different experimental conditions are compared. 
If these conditions simply differ in delay, this will lead to different recovered effect sizes and could thus be falsely interpreted as differences in directed dependency.

Further, the results reiterate that TE\textsubscript{SPO}, while featuring a constant effect across different ground truth interaction delays, recovers a flawed estimate of the interaction delay. 
DI\textsubscript{ce} on the other hand behaves favourably by returning a constant effect size at the correct interaction delay. 
However, due to its drastically reduced effect size in this band-limited interaction, its estimates are noisier.

\subsection*{Varying the signal-to-noise ratio in source and target signals}

\begin{figure}[!ht]
\centering
\includegraphics[width=.75\linewidth, trim={.5cm 1.1cm 1.9cm .8cm},clip]{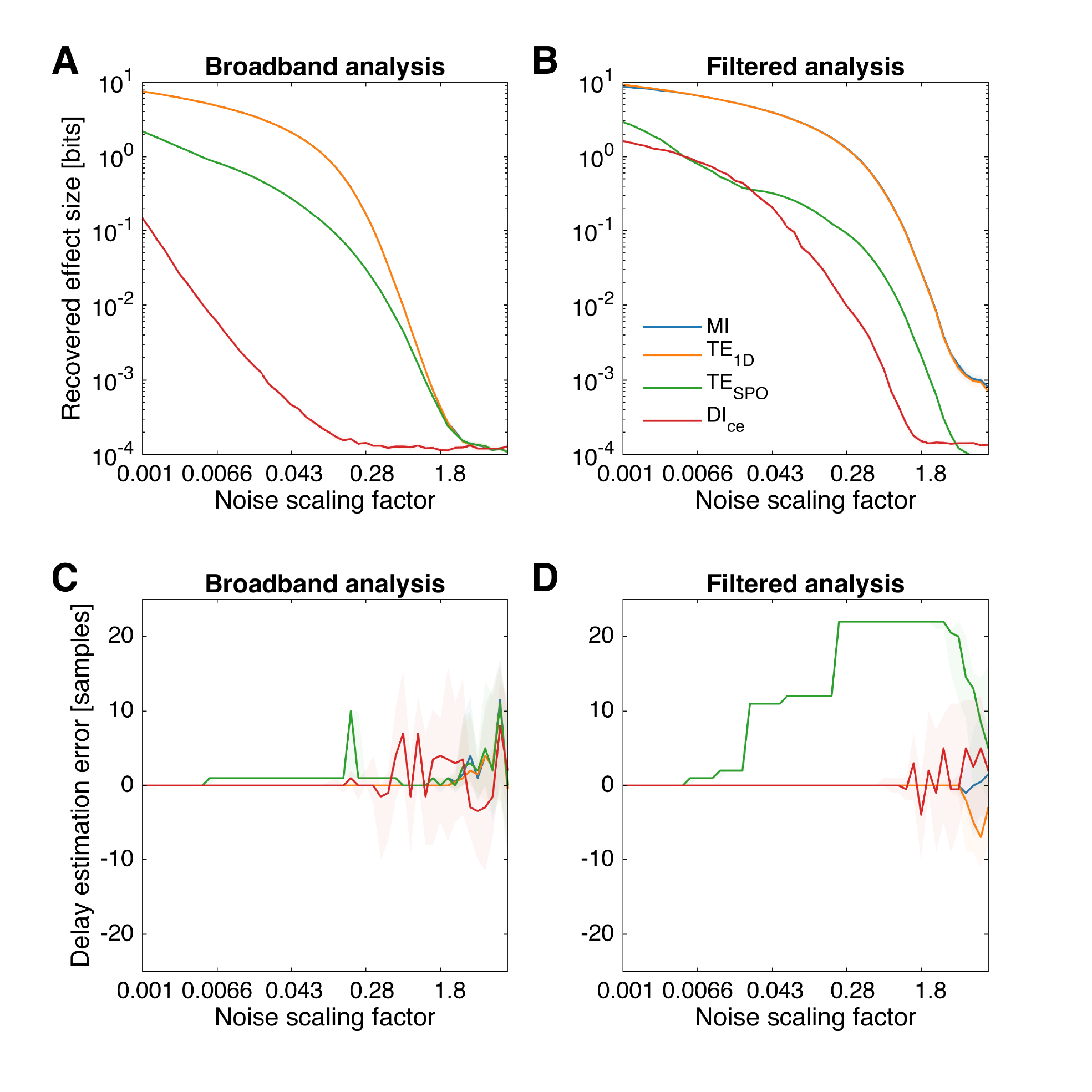}
\caption[Simulation with varying signal-to-noise ratios.]{\textbf{Simulation with varying signal-to-noise ratios.}

Basic setup of the simulation is the same as in \hyperref[fig_te_overview]{figure \ref{fig_te_overview}},  however, here, the amplitude of noise added to source and target varies (but the same amount is added to source and target), and 100 repetitions are sampled. 
Plots show the median across 100 repetitions, shaded regions denote bootstrapped 95\% confidence intervals. 
\textbf{A} In a broadband analysis scenario, all measures recover decreasing effect sizes with increasing noise amplitudes. 
DI\textsubscript{ce} recovers the smallest effect sizes and reaches the noise floor the earliest. 
\textbf{B} In a filtered analysis scenario, the recovered effect sizes are in general higher for all measures. 
\textbf{C} The recovered delays deteriorate as the noise increases. 
In a broadband analysis scenario, all measures correctly recover the ground truth interaction delay at low noise amplitudes. 
\textbf{D} In a filtered analysis scenario, TE\textsubscript{SPO} has higher delay reconstruction errors, while the delays recovered by the other measures benefit from the increased effect size.
}
\label{fig_te_noise}
\end{figure}

In a second simulation, we were interested in comparing the four measures when faced with increasingly noisy signals. 
While the ideal measure should be highly sensitive to a present effect and recover it at the correct delay, ignoring self-predictable parts of dependencies should inevitably reduce the effect size.

We indeed found that DI\textsubscript{ce} recovered the smallest effects and hit the noise floor at the lowest noise amplitude in comparison to the other measures (\hyperref[fig_te_noise]{figure \ref{fig_te_noise}}).
In a filtered analysis scenario, the recovered effects were in general higher than those in a broadband analysis scenario. 
Interestingly, for some noise amplitudes, DI\textsubscript{ce} recovered stronger effects than TE\textsubscript{SPO}. 
At very low noise amplitudes, all measures succeeded in recovering the correct delay irrespective of whether filters were applied in the analysis or not. 
However, as noise increased, especially TE\textsubscript{SPO} failed in recovering correct delay estimates. 
While this was especially problematic for filtered analyses, other measures had lower delay estimation errors at higher noise amplitudes due to the gain in sensitivity afforded by filtering.

\subsection*{Varying the signal-to-noise ratio independently in source and target signals}

In a third step, we were interested in assessing how the four measures would react to asymmetric variations of the SNR \citep{bastos2016}. 
The ideal measure of TE should recover an invariant interaction delay and an effect size that symmetrically decreases as noise of increasing amplitude is added to either source or target signals.

\begin{figure}[!ht]
\centering
\includegraphics[width=\linewidth, trim={3.2cm 3.4cm 2.3cm 2.25cm},clip]{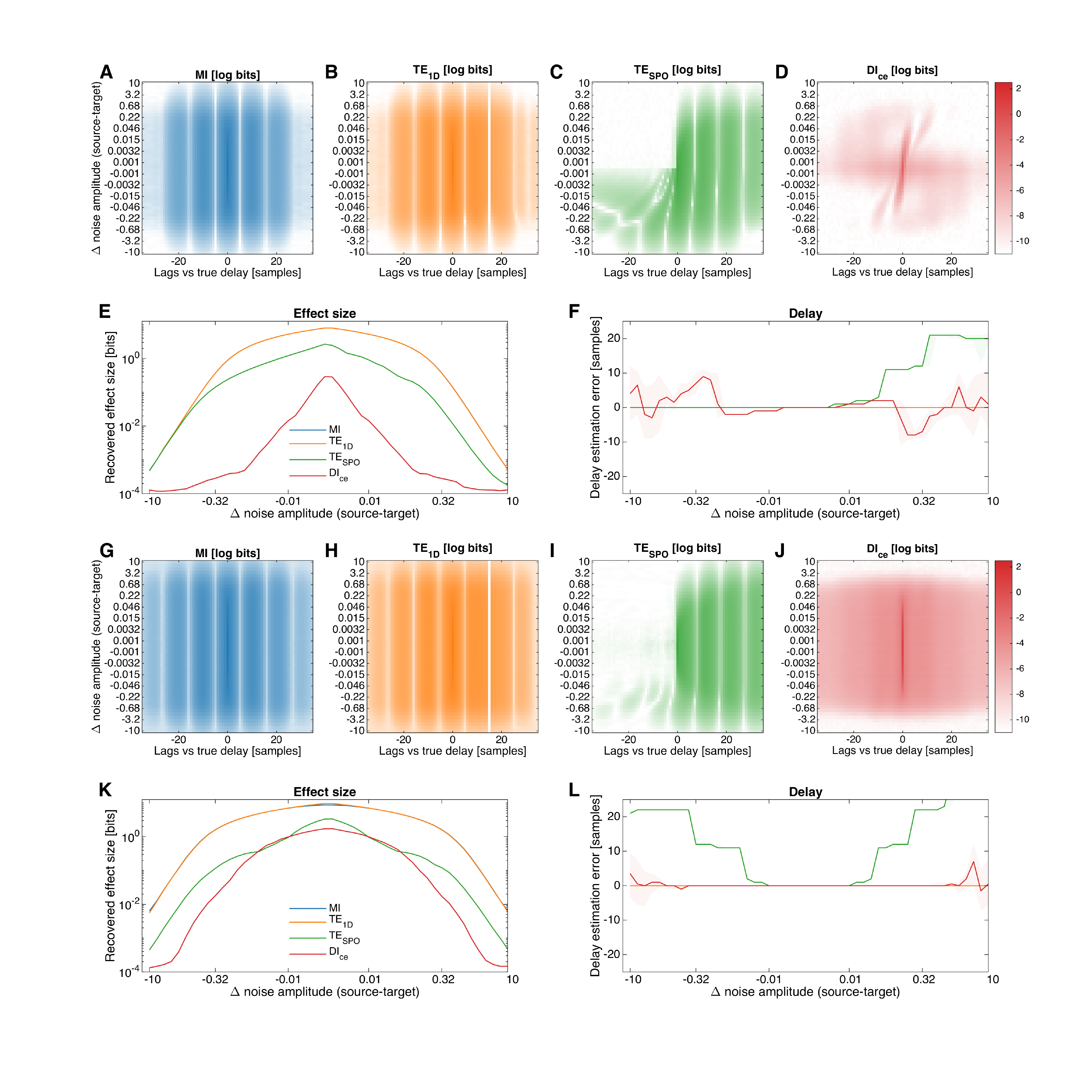}

\caption[Simulation with signal-to-noise ratios varying independently in source and target signals.]{\textbf{Simulation with signal-to-noise ratios varying independently in source and target signals.}

Basic setup of the simulation is the same as in \hyperref[fig_te_overview]{figure \ref{fig_te_overview}}, however, here, the amount of noise added to source or target signals is varied, and 100 repetitions are sampled. 
\textbf{A - D} show single trial delay profiles in a broadband analysis scenario for various measures, corrected for the true interaction delay across different noise amplitudes (negative delta on the y-axis refers to higher noise amplitude in the target signal, positive delta on the y-axis refers to a higher noise amplitude in the source signal).
The ideal measure should always return the highest values at a lag of 0 relative to the true delay. 
Only MI and TE\textsubscript{1D} succeed (but do not quantify Granger-causal information). 
\textbf{E} and \textbf{F} show the same as A -- D, but repeated for 100 trials.
Plots show the median across trials, shaded regions denote bootstrapped 95\% confidence intervals. 
The maximum effect size across lags for all measures symmetrically decays as the noise increases in either source or target signals. 
The delay misestimation however behaves asymmetrically, especially for TE\textsubscript{SPO}, which recovers wrong delays when the source is noisier than the target. 
\textbf{G -- J} The same as A -- D, but within a filtered analysis scenario. 
Here, DI\textsubscript{ce} performs favourably. 
\textbf{K} and \textbf{L} show the same as E and F, but within a filtered analysis scenario.
}
\label{fig_te_noiseAsym}
\end{figure}

\hyperref[fig_te_noiseAsym]{Figure \ref{fig_te_noiseAsym}} shows how the set of four measures behaved.
All measures succeeded in returning symmetric decreases of the recovered effect sizes when there was more noise in either source or target signals, irrespective of whether analysis filters were applied or not. 
However, only the undirected delayed MI as well as the practically undirected TE\textsubscript{1D} (see \hyperref[fig_te_overview]{figure \ref{fig_te_overview}}) recovered interaction delays that were unaffected in these settings. 
For TE\textsubscript{SPO}, an increasing SNR imbalance with noisier source signals led to stronger biases in the estimated interaction delay for broadband analyses. 
For analyses where filters were applied, these biases grew stronger for both noisier source and noisier target signals. 
DI\textsubscript{ce} exhibited biases in the broadband analyses, but performed favourably when analyses filters were applied.
This analysis additionally corroborates an argument of caution when interpreting TE results \citep{wollstadt2017}: when there are asymmetric changes in e.g. measurement noise across conditions, this does not automatically imply a change of coupling across these conditions, but might merely reflect the sensitivity of TE to such measurement noise.

\subsection*{Varying the bandwidths of the simulated effect as well as of the analysis filter}

In our fourth simulation, we reasoned that a characteristic effect which a suitable measure of TE should show is a varying recovered effect size as a function of the bandwidth of the ground truth effect. 
Specifically, the theoretical extreme case of a sinusoidal source signal and its phase-shifted copy as a target signal should lead to zero TE. 
Such signals have no degrees of freedom and are perfectly predictable from themselves. 
Additionally considering potential source signals can thus not add any information. 
As one turns to signals with increasingly broad passbands, these degrees of freedom increase. 
Consequently, a higher directed effect should be quantifiable. 

\begin{figure}[!ht]
\centering
\includegraphics[width=\linewidth, trim={2.25cm 1.1cm 2.7cm .8cm},clip]{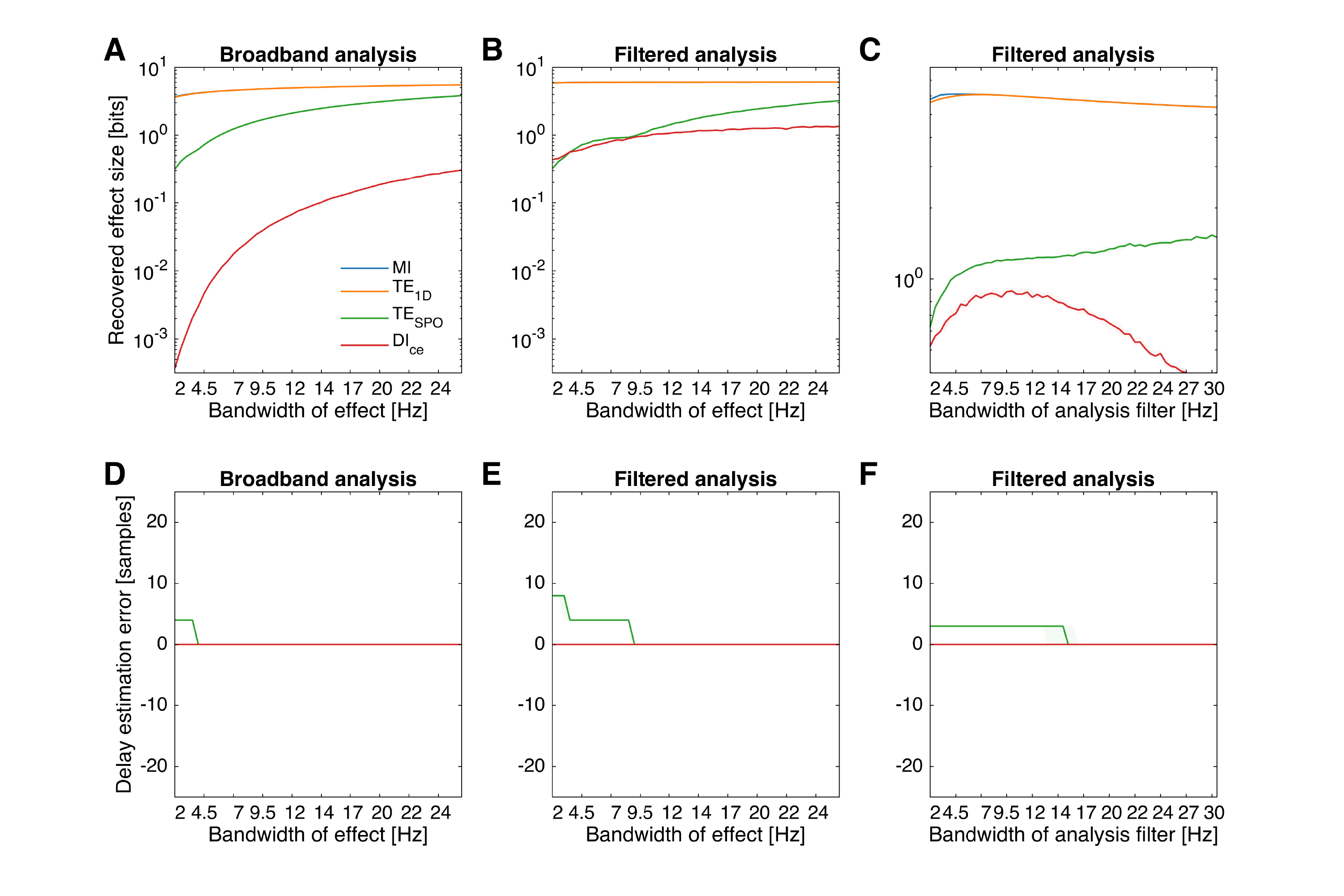}
\caption[Simulation with varying bandwidths of the transmitted signal.]{\textbf{Simulation with varying bandwidths of the transmitted signal.}

Basic setup of the simulation is the same as in \hyperref[fig_te_overview]{figure \ref{fig_te_overview}}, however, here, the ground truth bandwidth of the effect is varied, and 100 repetitions are sampled. 
Plots show the median across 100 repetitions, shaded regions denote bootstrapped 95\% confidence intervals. 
\textbf{A} The recovered effect size increases as the bandwidth of the effect grows. 
This increase however spans the most orders of magnitude for DI\textsubscript{ce}. 
\textbf{B} In a filtered analysis scenario, the effect sizes are stronger than in a broadband analysis scenario. 
\textbf{C} When varying the bandwidth of the filter used in the analysis, DI\textsubscript{ce} returns the peak effect size when the analysis filter matches the ground truth effect. 
All other measures fail to do so, TE\textsubscript{SPO} even increases monotonically as the analysis filter passband grows. 
\textbf{D - F} report the interaction delays recovered in the same simulations as shown in A - C. 
The previously reported misestimation of the interaction delay of TE\textsubscript{SPO} is limited to narrow band effects.
}
\label{fig_te_bandwidth}
\end{figure}

When testing this with the set of four measures (\hyperref[fig_te_bandwidth]{figure \ref{fig_te_bandwidth}}), we found that all measures monotonically increased in effect size as the bandwidth of the effect increased.
However, in a broadband analysis, this increase spanned 2.5 orders of magnitude for DI\textsubscript{ce}, which thus had the strongest relationship between its recovered effect size and increased degrees of freedom of the input signals. 
This was however also driven by the ground truth effect covering a larger part of the entire frequency spectrum. 
In a filtered analysis scenario where filter parameters were chosen to isolate the effect from the noise, only TE\textsubscript{SPO} and DI\textsubscript{ce} showed increases in effect size with broader effect bandwidths.
The recovered delays were, as shown in previous analyses, unaffected for all measures except for TE\textsubscript{SPO}.
The latter exhibited stronger biases in the recovered delay for narrower simulated effect bandwidths.

In theory, applying analysis filters that match the ground truth effect should return a stronger effect size than analysis filters that are too broad or too narrow. 
Too broad analysis filters should fail to reduce the impact of noise in neighbouring frequencies, while too narrow analysis filters should ignore variance of the effect of interest and thus reduce the sensitivity. 
When applying analysis filters of varying width to a simulated coupled system of fixed ground truth effect bandwidth, DI\textsubscript{ce} showed the clearest peak close to the ground truth effect (bandwidth of 10 Hz, \hyperref[fig_te_bandwidth]{figure \ref{fig_te_bandwidth}}C) across analysis filter widths. 
TE\textsubscript{SPO} on the other hand exhibited a monotonic increase in effect size as the analysis filter was broadened, which would thus result in mis-estimations of the interaction bandwidth. 
The undirected delayed MI and the practically undirected TE\textsubscript{1D} only showed comparably weak variance across different analysis filters.

\subsection*{Varying the centre frequency of the simulated effect}

In a last simulation, we were interested in assessing the stability of the set of four measures when the centre frequency was varied. 
A suitable measure of TE should recover effect sizes and interaction delays with no variation under different ground truth centre frequencies, and it should find the strongest effects at the ground truth frequency.

\begin{figure}[!ht]
\centering
\includegraphics[width=\linewidth, trim={3cm 1.25cm 3.4cm .5cm},clip]{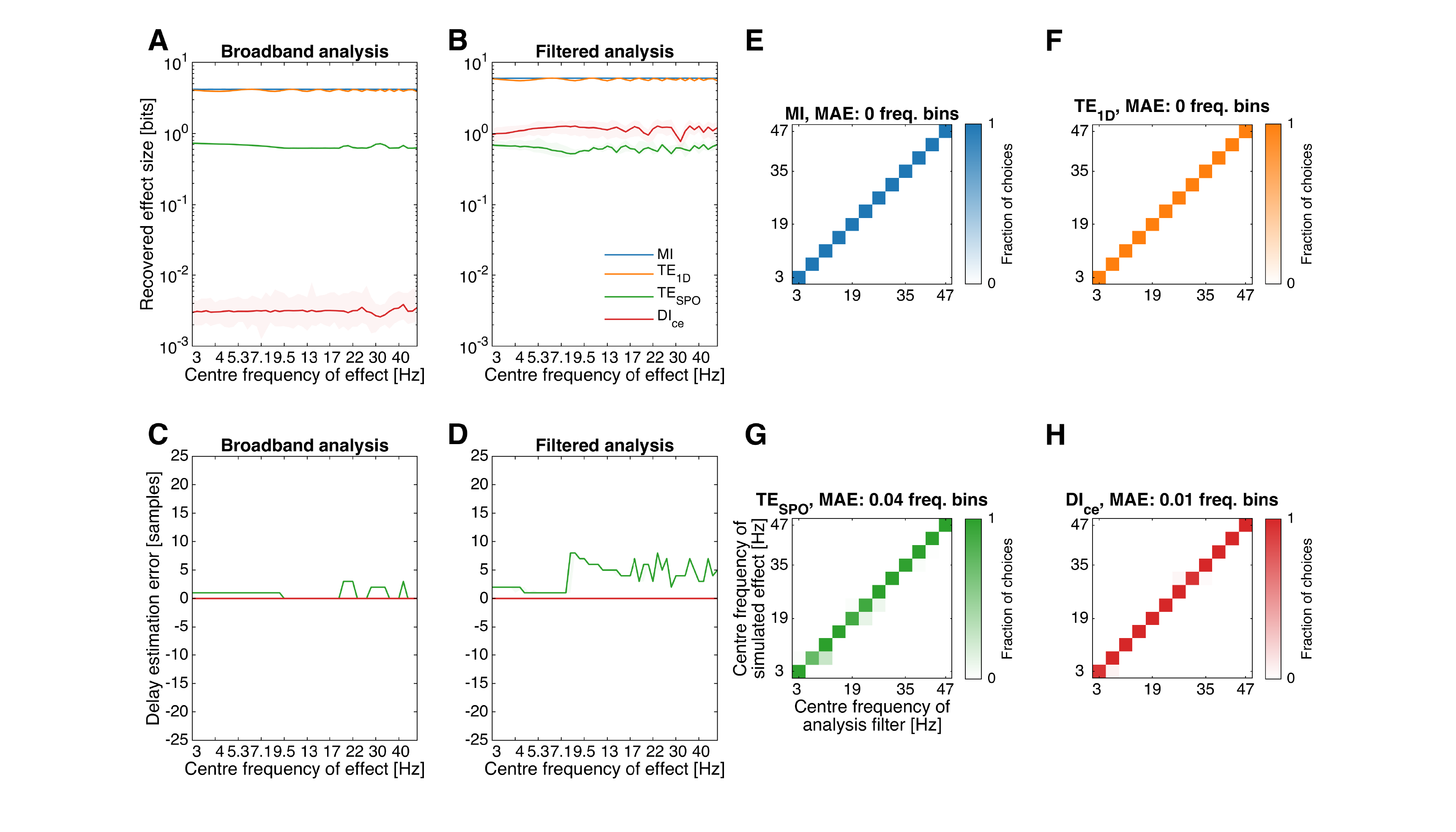}
\caption[Simulation with varying centre frequencies of the effect.]{\textbf{Simulation with varying centre frequencies of the effect.}

Basic setup of the simulation is the same as in \hyperref[fig_te_overview]{figure \ref{fig_te_overview}}, however, here, the centre frequency of the effect is varied, and 100 repetitions are sampled. 
Line plots show median across 100 repetitions, shaded regions denote bootstrapped 95\% confidence intervals. 
\textbf{A} In a broadband analysis scenario, all measures recover effect sizes that are highly constant across centre frequencies. 
TE\textsubscript{1D} however exhibits a slight ringing, TE\textsubscript{SPO} and DI\textsubscript{ce} recover noisier estimates. 
\textbf{B} The same holds for a filtered analysis scenario, where DI\textsubscript{ce} and TE\textsubscript{SPO} exhibit stronger variations. 
\textbf{C} and \textbf{D} As shown in other simulations, TE\textsubscript{SPO} suffers from delay estimation problems. 
These vary especially strong across centre frequencies for filtered analysis scenarios. 
\textbf{E - H} Confusion matrices obtained when scanning data with different ground truth interaction frequencies with a bank of analysis filters (choosing the frequency with the maximum across analysis filters). 
All measures generally succeed in returning peak effects at the true interaction frequency. 
TE\textsubscript{SPO} has the highest error. 
}
\label{fig_te_centreFreq}
\end{figure}

We found that all measures considered here recovered effect sizes that were indeed relatively stable across different ground truth centre frequencies in both broadband and filtered analysis scenarios (\hyperref[fig_te_centreFreq]{figure \ref{fig_te_centreFreq}}).
TE\textsubscript{1D} and DI\textsubscript{ce} however did exhibit slight biases, especially for the combination of DI\textsubscript{ce} and higher simulated centre frequencies. 
All measures except TE\textsubscript{SPO} recovered unbiased estimates of the interaction delay for all tested centre frequencies.
TE\textsubscript{SPO} on the other hand exhibited a marked variation of the recovered interaction delay across the different centre frequencies.

Finally, when analysing the simulated data with filters across different bands, all measures mostly peaked when analysis filters matched the ground truth frequency. 
For MI and TE\textsubscript{1D}, the mean absolute error (MAE) across frequency bands was 0 frequency bins, for DI\textsubscript{ce} it was .01 frequency bins and for TE\textsubscript{SPO}, the MAE was strongest with .04 frequency bins. 
Identifying the correct spectral band thus did not seem to pose a particular problem for any of the considered measures.

\subsection*{Studying low-frequency MEG speech envelope tracking with directed measures}

Finally, we wanted to explore the behaviour of our measure on real data. 
To do so, we turned to a dataset of $n=24$ participants who listened to a continuous narrative of $1$ hour duration while their MEG was recorded \citep{daube_simple_2019}.
Examples of cerebro-peripheral coupling such as this present a good testbed for frequency-specific measures of directed connectivity, because the ground truth direction of the effect is known. 
Further, a plethora of studies has examined the relationship of MEEG responses to the time varying energy, or “amplitude envelope”, of the speech stimulus in the delta and theta bands \citep{ding2014}.
However, it is usually studied using bivariate, undirected measures of connectivity that do not consider the self-predictability of the MEG responses. 
The degree to which such bivariate dependencies could be accounted for by auto-regressive models of the neuronal response signal could in principle reflect the degree to which bivariate speech tracking is a signature of predictive rather than reactive bottom-up processing. 
On the other hand, effects found by directed measures of dependency (under the assumption of a given auto-regressive model) would be stronger evidence of reactive, bottom-up processing of unpredictable parts of the stimulus input.

\begin{figure}[!ht]
\centering
\includegraphics[width=\linewidth, trim={3cm 3.8cm 3.4cm 2.2cm},clip]{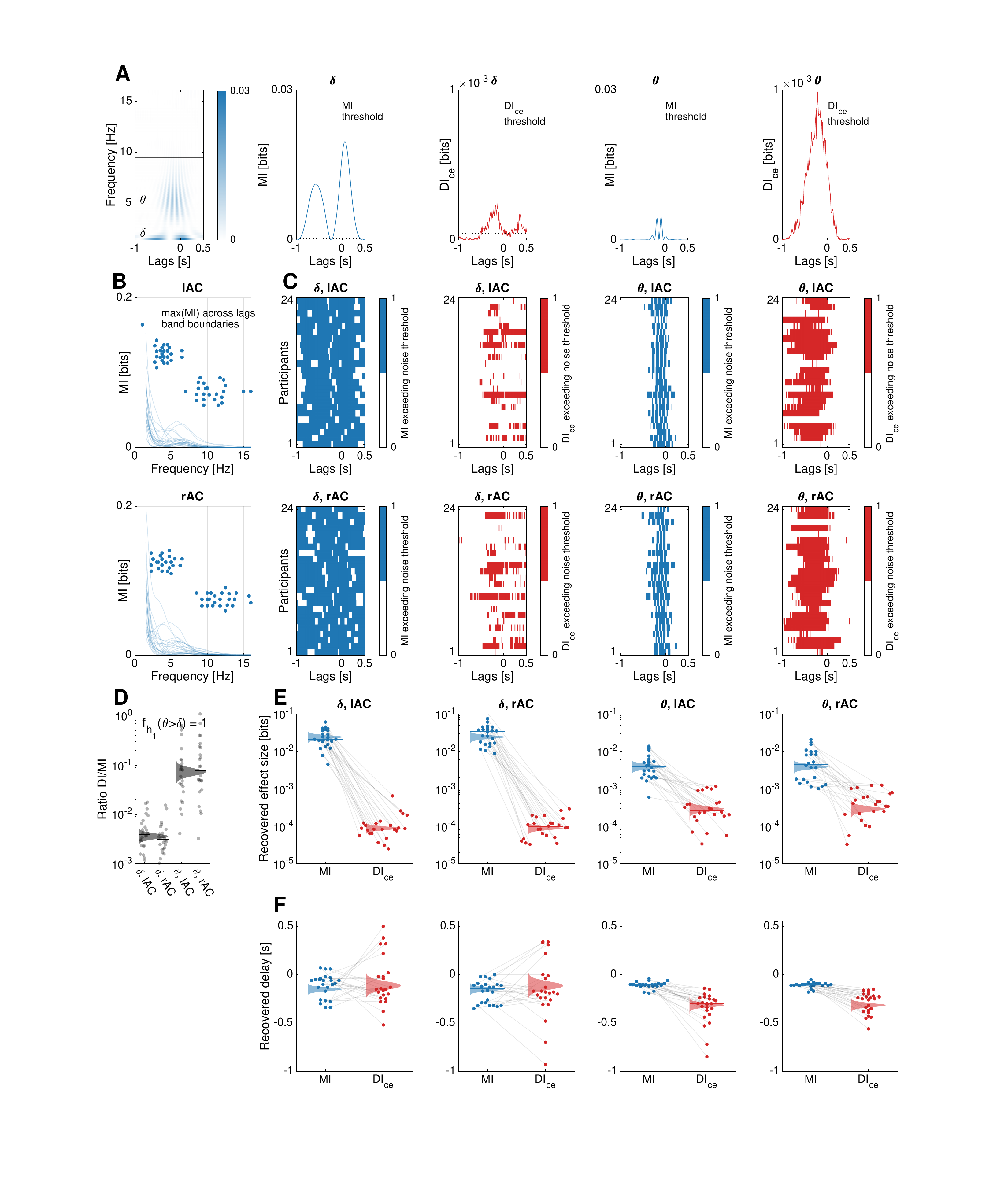}
\captionsetup{list=no}
\caption{Caption on following page.}
\end{figure}
\begin{figure}[!ht]
\captionsetup{labelformat=adja-page, list=yes}
\ContinuedFloat
\caption[Results obtained from source level MEG recordings (left and right auditory cortices, l and r ACs) obtained during continuous speech listening (n=24).]{\textbf{Results obtained from source level MEG recordings (left and right auditory cortices, l and r ACs) obtained during continuous speech listening (n=24).}

\textbf{A} Results for a typical participant. 
Plots show spectrotemporally resolved delayed MI with individual boundaries of delta and theta bands overlaid as black horizontal lines (leftmost plot) as well as delay profile of delayed MI and DI\textsubscript{ce} in delta and theta bands, with noise thresholds overlaid as black dotted lines.
\textbf{B} Spectral profiles of delayed MI for all participants in left (top) and right (bottom) ACs. 
Points denote upper pass-band cut-off frequencies of the delta (upper scatter) and theta (lower scatter) bands. 
\textbf{C} Binarised delayed MI as well as 
DI\textsubscript{ce} delay profiles for all participants in delta and theta bands in left and right ACs. 
While delayed MI in both bands and DI\textsubscript{ce} in the theta bands cross the noise threshold in most cases, DI\textsubscript{ce} in the delta band is often close to or below the noise threshold. 
\textbf{D} Ratios of DI/MI in delta and theta bands of left and right ACs for each individual participant, medians overlaid as black lines, density estimates from posterior distributions overlaid as transparent shapes. 
Theta band bivariate tracking can be less explained by auto-prediction of the MEG signal than delta band tracking. 
\textbf{E} Effect sizes underlying the computation of the ratio in D in delta and theta bands of the left and right ACs in each individual participant. 
Medians are overlaid as coloured lines, density estimates from posterior distributions overlaid as transparent shapes. \textbf{F} Delays recovered by delayed MI and DI\textsubscript{ce} in the delta and theta bands of left and right ACs. 
Since DI\textsubscript{ce} fails to cross the noise threshold in many cases for the delta band, these delay estimates are uninterpretable. 
In the theta band, DI\textsubscript{ce} recovers a longer delay than delayed MI.
Also see Figures \hyperref[fig_te_sup_passbands]{\ref{fig_te_sup_passbands}} -- \hyperref[fig_te_sup_pp]{\ref{fig_te_sup_pp}}.
}
\label{fig_te_meg}
\end{figure}

\begin{figure}[!ht]
\centering
\includegraphics[width=\linewidth, trim={3cm 3.4cm 3cm 2.5cm},clip]{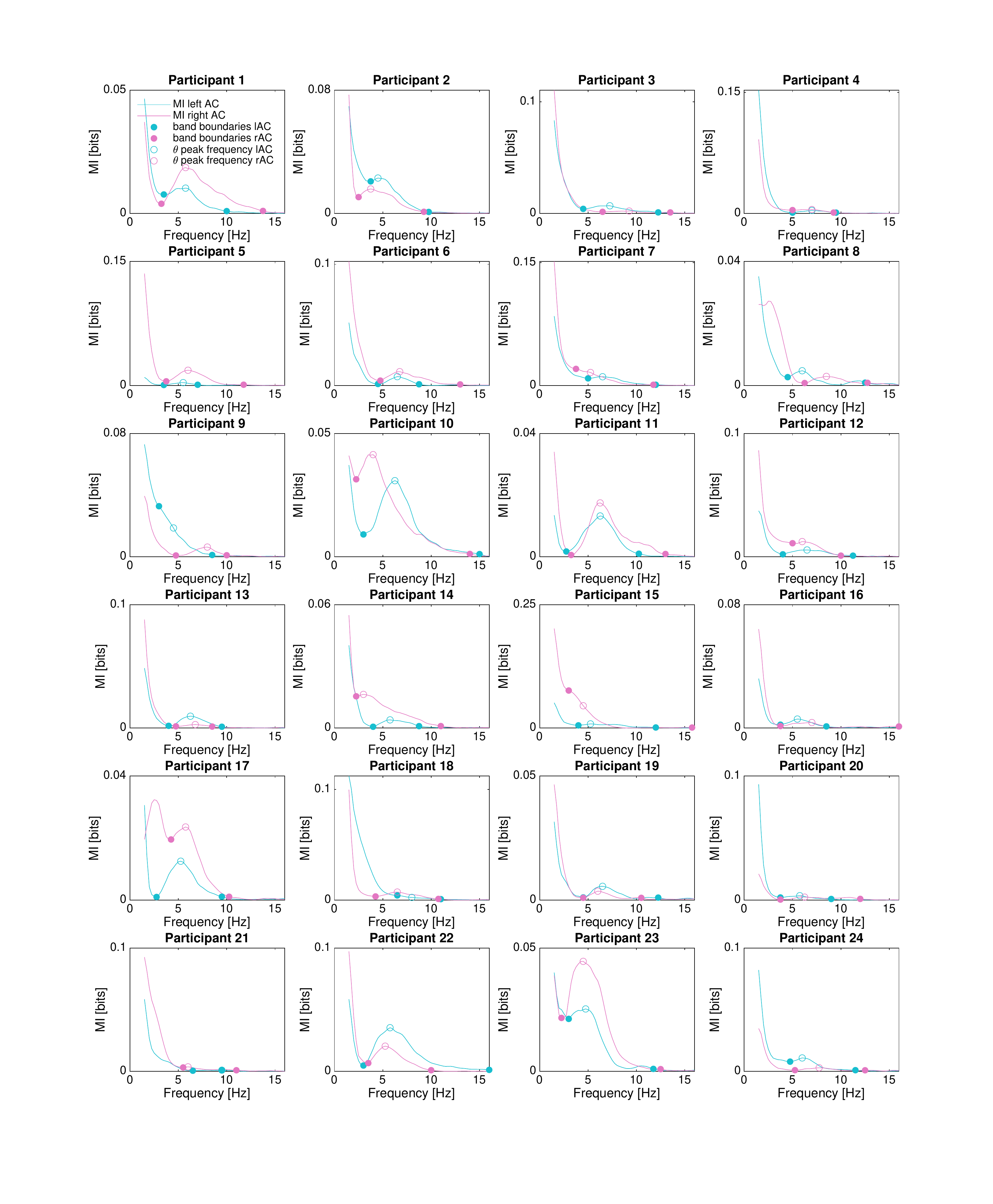}
\caption[Spectra of MI for left and right auditory cortices for each individual participant.]{\textbf{Spectra of MI for left and right auditory cortices for each individual participant (related to \hyperref[fig_te_meg]{figure \ref{fig_te_meg}}).}

Individual boundaries of delta and theta bands are overlaid as filled circles, theta peak frequencies (used for \hyperref[fig_te_sup_megFixedFreq]{figure \ref{fig_te_sup_megFixedFreq}}) are overlaid as empty circles.
}
\label{fig_te_sup_passbands}
\end{figure}

In most participants, we found spectra of delayed MI (see \hyperref[fig_te_meg]{figure \ref{fig_te_meg}B} and \hyperref[fig_te_sup_passbands]{figure \ref{fig_te_sup_passbands}}) that were suggestive of two spectral components involved in speech tracking which we will refer to as delta and theta bands \citep[note that our functional definition of the theta band had higher upper cutoffs than the canonical theta band, which is commonly defined between 4 to 8 Hz, ][]{klimesch1999, wang2010}.
We found super-threshold delayed MI in both left and right auditory cortices of all participants in both delta and theta bands (\hyperref[fig_te_meg]{figure \ref{fig_te_meg}C}). 
This was stronger in delta than in theta frequency bands (fraction of samples in favour of hypothesis $f_{h_1} = 1$, \hyperref[fig_te_meg]{figure \ref{fig_te_meg}E}, see also \hyperref[fig_te_sup_ppp]{figure \ref{fig_te_sup_pp}}). 
With DI\textsubscript{ce} however, we found only weak effects that barely exceeded the noise thresholds in the delta bands, while effects in the theta bands exceeded noise thresholds with only 1 exception in the left and 2 exceptions in the right hemisphere (\hyperref[fig_te_meg]{figure \ref{fig_te_meg}C}).
This constitutes strong evidence for a robust population-level prevalence of the effect \citep{ince_bayesian_2021}.
These theta DI\textsubscript{ce} effects were generally much stronger than in the delta band ($f_{h_1} = 1$; \hyperref[fig_te_meg]{figure \ref{fig_te_meg}E}, see also \hyperref[fig_te_meg]{figure \ref{fig_te_meg}E}, see also \hyperref[fig_te_sup_ppp]{figure \ref{fig_te_sup_pp}}). 
These results translated into a difference of ratios of DI divided by MI between the delta and theta frequency bands, where we found higher ratios in the theta than in the delta band in both left and right auditory cortices ($f_{h_1} = 1$; \hyperref[fig_te_meg]{figure \ref{fig_te_meg}D}, see also \hyperref[fig_te_meg]{figure \ref{fig_te_meg}E}, see also \hyperref[fig_te_sup_ppp]{figure \ref{fig_te_sup_pp}}). 
This suggests that the bivariate speech tracking in the theta band could consist of more bottom-up and reactive processing than the bivariate speech tracking in the delta band.
We had found in simulation analyses that the effect size recovered by DI\textsubscript{ce} grows as a function of increasing bandwidth (see \hyperref[fig_te_bandwidth]{figure \ref{fig_te_bandwidth}}). 
Since our individualised definitions of delta and theta bands led to wider theta than delta analysis filters, we repeated this analysis while constraining the analysis filter bandwidth to be the same in delta and theta bands. 
We found that the difference in DI over MI ratios of the delta vs the theta bands shrunk, but persisted under this constraint ($f_{h_1} = 1$; see figures \hyperref[fig_te_sup_megFixedFreq]{\ref{fig_te_sup_megFixedFreq}} and \hyperref[fig_te_sup_pp]{\ref{fig_te_sup_pp}}). 
We could thus rule out that analysis bandwidth alone could explain the difference in DI/MI ratios in delta vs theta bands. 

Given the overall relatively narrow passbands of both delta and theta bands, the measurable directed effects were in general up to several orders of magnitude lower than the undirected effects, suggesting that predictive processing could indeed make up the lion’s share of delta and theta envelope tracking.

Lastly, we considered the delays recovered by delayed MI and DI\textsubscript{ce} in delta and theta bands (\hyperref[fig_te_meg]{figure \ref{fig_te_meg}F}).
For the delta band, the delays recovered by DI\textsubscript{ce} spanned a wide range, even reaching into regions suggestive of the MEG signal preceding the speech signal for some participants. 
However, since the effect sizes in the delta band were in many cases close to or below the noise threshold, these recovered delays were uninterpretable. 
For the theta band however, effect sizes were robust in most cases. 
Interestingly, the delays recovered by DI\textsubscript{ce} (lAC: median across participants of -.43s, rAC: median of -.40s) were longer than those recovered by delayed MI (lAC: median of -.10s, rAC: median of -.10s; $f_{h_1} = 1$), suggesting a slower bottom-up, reactive processing. 

Taken together, we take our results as aligning with models that characterise the lion’s share of activity of auditory cortices visible in MEG at a given point in time as reflecting predictions of the stimulus at a short delay. 
Such predictions can be formed on the basis of an integration of reactive, bottom-up processing of stimulus input at longer delays with an internal language model.

\section*{Discussion}

In this study, we have addressed the problem of quantifying band-limited directed interactions in bivariate sets of source and target signals.
With a simplistic and intuitive simulation setup, we have highlighted shortcomings of common estimators of TE when facing this problem. 
Our proposed alternative DI\textsubscript{ce}, relying on a type of temporal whitening of the source and target signals, overcomes these shortcomings. 
With an array of simulations, we extensively characterised DI\textsubscript{ce} in relation to undirected MI as well as common estimators of TE. 
Lastly, we turned to a dataset of continuous speech listening in MEG to study the speech envelope tracking specific to delta and theta bands from the perspective of a directed measure. 
We found that in such narrow bands with essentially low degrees of freedom, the measurable directed effects are very small in comparison to undirected effects. 
Moreover, with DI\textsubscript{ce}, we found that theta speech tracking has a stronger directed part than delta speech tracking, and that the directed theta speech tracking has a longer delay and a broader temporal profile than the undirected tracking. 
The directed effects could potentially reflect a purely bottom-up processing stage at a later delay. 
On the basis of this reactive processing, the auditory cortical system could then generate predictions about the upcoming speech acoustics, which the undirected effects could be a signature of.

Many of the problems and perspectives highlighted in this study are not new to the literature.
We are for example not the first to point out problems of TE estimators relying on only a single sample of the target past signal at a delay equivalent to that of the source to the target at a given scanned interaction delay \citep{wibral2013}. 
However, we hope that our intuitive demonstrations contribute to a more widespread appreciation of the high similarity of the resulting “TE”\textsubscript{1D} and simple delayed MI as well as variations of recovered effect sizes in the presence of different ground truth interaction delays.
Likewise, the delay misestimation problems of more sophisticated estimators such as TE\textsubscript{SPO} when faced with narrow-band effects have been reported before \citep{wollstadt2017}. 
In the same way, the PID perspective on conditional MI based TE and the implication of including synergistic interactions has been developed previously \citep{james2016}. 
Here, we found that from this PID perspective on the TE problem, the delay mis-estimations of TE\textsubscript{SPO} could mostly be attributed to such synergistic interactions of the source and the target that TE estimators relying on conditional MI pick up on. 

An intuitive solution to fix this would be to instead consider the unique information of the source about the target, and thus effectively ignore the synergistic contributions included in conditional MI \citep{barrett2015}.
However, we found that this was neither particularly sensitive in detecting effects nor accurate in recovering the ground truth delays, albeit with much less error than synergy-driven conditional MI based estimators. 
Our proposed estimator, DI\textsubscript{ce}, circumnavigates these issues by not relying on conditional MI in the first place, but by instead converting the source and target time series into time series of sample wise conditional entropy. 
This effectively frees the signals from predictable parts, ensuring that a subsequently computed delayed MI cannot be affected by autocorrelation. 
Applying this whitening to the target signal only would be analogous to the basic idea of TE, but would introduce asymmetric temporal distortions in the target relative to the source, resulting in erroneous delay estimates. 
Again, the general idea of such a two-staged approach is not new as such \citep{haugh1976, cliff2021}. 
However, we have here developed it from a simple cross-correlation of residuals \citep{haugh1976}, which only takes into account the mean of a predicted time series, to the delayed MI of sample wise conditional entropy, which instead relates a given sample to the mean and variance of a predicted distribution. 
On the basis of our simulations, we put forward that it does offer a useful perspective on the problem of estimating directed interactions of band-limited processes at the correct delay.

In the light of various careful considerations of the pitfalls associated with this goal \citep{florin2010, barnett2011}, we thus hope that we can contribute to a more widespread adoption of directed perspectives on band-limited dependencies.
Within neuroscience, such genuine narrow-band signal components are arguably not in all places where aperiodic signals have been straight-jacketed into oscillations by means of band-pass filters \citep{donoghue2020, gerster2021}.
In fact, depending on the algorithm used to define oscillatory components, even the delta component identified by us is potentially attributable to the aperiodic part of the MEG spectrum. 
In general, oscillatory components are nevertheless indisputably widespread \citep{wang2010}. 
It is in such situations of high auto-correlation where the consideration of self-predictability should have the most obvious and strongest effects, and where the use of measures like TE is thus of potentially high interest. 
After all, frequency-specific oscillatory components are at the heart of popular conceptualisations of interactions of neuronal populations \citep{schnitzler2005, fries2015}. 
We conjecture that estimators of directed interactions that return counterintuitive results in such circumstances as simplistically simulated may not be well matched to the interpretations typically applied. 
This problem has been recognised elsewhere, and \citep{pinzuti2020} propose a clever permutation scheme to address this issue. 
Here, we instead suggest to prioritise the interpretability of the directly measured effect size.

In this spirit, our measure is in principle a versatile tool applicable to arbitrary (neuro-)scientific questions. 
We however subscribe to the view that as such, mechanistic interpretations of cerebro-cerebral dependencies are prohibitively hard in many cases \citep{mehler2018}. 
Given the highly incomplete picture of the entire neuronal activity that is accessible with any given neuroimaging modality, it is virtually impossible to rule out that a given dependency stems from a third unobservable region affecting the supposed source and target. 
Directed dependency measures per se thus never warrant the inference of causality. 
This problem is however alleviated in cerebro-peripheral settings, where the dependency of neuroimaging signals on external signals is studied \citep{gross2021}. 

The question of causality is thus arguably less controversial in our example case of the dependence of MEG signals on a continuous acoustic speech stimulus. 
According to the burgeoning field of predictive coding however, the brain appears to predict the upcoming speech input \citep{brodbeck2018b, donhauser2020, heilbron2021}, rendering temporal relations of time series of assumed cause and effect less trivial. 
Our suggestions regarding these problems are mostly of an indirect nature, arguing that the part of undirected dependencies that directed measures deliberately ignore could reflect predictive processing, while what directed measures quantify should reflect reactive, bottom-up processing. 
It is important to note however that simply because the part of a neural response that we can observe with a given neuroimaging modality is predictable from its own past does not automatically imply that it is in fact being predicted by the brain \citep{dewit2016}. 
Response components identified with impulse response functions as measured with autocorrelations, temporal response functions \citep{crosse2016} or here delayed MI have been suggested to be analogous to the well-studied components of auditory event-related potentials \citep{lalor2009}. 
These do occur in response to unpredictable events \citep{ritter1968}, implying that the underlying generators are not at all restricted to prediction. 
It has however also been suggested that the low-frequency potential “entrains” in response to sustained stimuli such as continuous speech, especially when they are (quasi-)rhythmic \citep{lakatos2008, lakatos2019}. 
Viewing ERPs in response to unpredictable events through this lens, they could manifest the launch or “reset” of such predictions for subsequent events \citep{sayers1974, makeig2002}. 
Such entrainment would serve to align phases of low and high neuronal excitability to relevant parts of the stimulus, such that crucial stimulus information could be encoded with high fidelity.
In this sense, the low-frequency potential is indeed often seen as a neuronal signature of a prediction about the upcoming stimulus. 
A further interesting detail of our results in this respect is the distribution of lags recovered by delayed MI (i.e. the timing of the peak of the delay profile). 
This was relatively broad in the delta frequency band, with the delays of some participants even suggesting the stimulus to follow the response, a hallmark of prediction that has been observed and interpreted similarly before in EEG delta-band responses to continuous speech \citep{etard2019}. 
Interestingly, such slow delta tracking signatures disappear when speech of a language foreign to the participant is listened to \citep{ding2016}, which is to be expected if it is seen as a predictive component relying on an internal language model. 

Under this perspective, the reactive, bottom-up part of the coupling carved out by DI\textsubscript{ce} would then essentially be driven by adjustments of the predictions in reaction to unpredictable parts in the stimulus. 
An important constraint of our proposed approach however follows from the choice of the model used to whiten the signals. 
Our choice of a non-uniform embedding to predict a signal’s present state from its own past is rooted in the history of TE, and its suitability to model an actual hypothesised biological acoustic model is disputable. 
In principle, there is nothing that prohibits the combination of our framework with more powerful autoregressive models such as recurrent neural networks. 
That the auditory system is actually entertaining correspondingly complex multi-level predictions of the upcoming input is an increasingly popular perspective \citep{donhauser2020, koskinen2020, heilbron2021, jain2021, schmitt2021, caucheteux2021b}. 
This would then effectively further shrink the effect sizes that our proposed DI\textsubscript{ce} could discover, assigning an even smaller portion of variance to unambiguously bottom-up reactive processing than found here.

Lastly, it is interesting to compare our approach to a recently increasingly popular analysis strategy in the field of predictive coding. 
This strategy usually builds on the framework of encoding models \citep{naselaris2011}, where first a so-called linearising feature space (a nonlinear transformation of the stimulus) is identified and then linearly mapped onto the brain response. 
For questions of predictive coding, researchers make use of powerful predictive models of a stimulus class to derive measures of surprisal (and uncertainty) associated with a given part of the stimulus given its preceding context to obtain the linearising feature space \citep{brodbeck2018b, donhauser2020, koskinen2020}. 
Our transformation of the signal of time varying energy of the speech stimulus into a time-series of conditional entropy can be seen as a simple version thereof. 
Within the standard approach, it is then however uncommon to perform a similar operation on the neuronal response, which is central to our proposed measure. 
The standard approach makes sense from the viewpoint of the hypothesis that the neuroimaging signal mainly reflects a prediction error. 
The conceptualisation of neuronal signals as reflecting prediction errors however decreases in its appeal to the degree to which the neuronal signal is predictable from itself. 
How could the brain be surprised if it already knew it was going to be surprised? 
We thus hypothesise that studies of prediction errors could increase their specificity by including a whitening of the target signals akin to what we suggest here. 
In theory, this should allow a more detailed characterisation of processes related to prediction errors. 

Taken together, this study makes a threefold contribution to the problem of quantifying band-limited directed dependencies: 
We offer simplistic yet intuitive and accessible simulations, propose our own estimator based on an information theoretic pre-whitening as well as provide an application to the case of speech tracking in MEG in delta and theta bands. 
With this, we hope to spark further interest in directed perspectives on frequency-specific interactions.

\section*{Methods}

\subsection*{Estimation of information theoretic quantities}
We used the Gaussian Copula Mutual Information framework \citep[gcmi,][]{ince2017b}. 
Here, the basic idea is to transform variables into standard normals to then apply closed-form expressions for information theoretic quantities of Gaussians. 
These can be derived as follows. 
The entropy $H$ of a variable $X$ is given by the expected value of the surprisal:

\begin{equation} \label{eq_te_entropy}
H(X)=E(h(x))
\end{equation}

For a Gaussian continuous variable $X$, this corresponds to the ``global” differential entropy:

\begin{equation} \label{eq_te_diffEntropy}
H(X)=\int_{-\infty}^{\infty} f(x) h(x) d x
\end{equation}

Here, the local or ``sample wise” contribution, i.e. the surprisal (also called ``information content”, ``Shannon information” or ``self-information”), is thus given by:

\begin{equation} \label{eq_te_localEntropy}
h(x)=-\log (f(x))
\end{equation}

For a Gaussian continuous variable of $k$ dimensions, this is equal to to the negative log-likelihood:

\begin{equation} \label{eq_te_nll}
h(x)=-\left(-\frac{1}{2}\left[\log (|\Sigma|)+(x-\mu)^{\prime} \Sigma^{-1}(x-\mu)+k * \log (2 \pi)\right]\right)
\end{equation}

$H(X)$ can then be simplified to:

\begin{equation} \label{eq_te_entropySimple}
H(X)=\frac{k}{2}+\frac{k}{2} \log (2 \pi)+\frac{1}{2} \log (|\Sigma|)
\end{equation}

We can then compute the mutual information between the variables $X$ and $Y$ (where $[]$ denotes concatenation to obtain a joint variable) follows:

\begin{equation} \label{eq_te_MI}
M I(X, Y)=H(X)+H(Y)-H([X Y])
\end{equation}

This pragmatic estimator comes at the cost of not being able to quantify all possible nonlinear effects, as information theoretic measures should do in theory.
Instead, it quantifies relationships described by a Gaussian copula. 
However, it avoids the loss of information and incompatibility with multidimensional variables as well as higher-order information theoretic quantities incurred by binning, a similarly pragmatic choice. 
Further, it is computationally cheaper and has a higher sensitivity for rank relationships than more sophisticated estimators such as nearest neighbour approaches, which can resolve nonlinear effects, but can in practice only be computed on vast amounts of samples from variables of relatively limited dimensionality.
In situations where nonlinear effects are of interest and resolvable when using a binned estimator with a realistically low amount of bins (such that each bin of the joint distribution is sufficiently sampled), such estimators are a pragmatic complement to gcmi.

\subsection*{Delayed mutual information}
The simplest information theoretic approach of estimating a dependency between source and target variables is computing the delayed mutual information (MI). 
It is the information theoretic equivalent of cross-correlation, where versions of the source- and target variables with a range of lags to each other are generated to then compute the dependency measure at each lag and obtain a delay profile of the relationship. 
In principle, the lagging of the source and target variables to each other can be implemented either way, by lagging the source and fixing the target or vice versa. 
To simplify concrete explanations that follow, we here choose an implementation where the source is lagged while the target is kept fixed. 
We thus define delayed MI between time series $X$ and $Y$ at an interaction delay $\delta_a$ as:

\begin{equation} \label{eq_te_delayedMI}
\text { delayed } M I(X, Y)_{\delta_{a}}=M I\left(X_{\delta_{a}}, Y_{0}\right)
\end{equation}

Here, $X_{\delta_{a}}$ denotes the lagged source with negative values of $\delta_a$ corresponding to delays where the source precedes the target and positive values corresponding to delays where the source follows the target. 
We trim the target present $Y_0$ to be of an equal number of samples as the lagged source.

\subsection*{Transfer entropy}
Originally, transfer entropy (TE) has been described as a conditional MI \citep{schreiber2000}. 
Instead of computing the simple delayed MI at each lag, the idea is to condition this MI on an operationalisation of the target past. 
In principle, the resulting quantity is thereby supposed to quantify to what extent the target signal can be explained by the source signal over and above to the extent to which the target signal can be explained by its own past.
This idea is equivalent to that of Granger causality \citep{granger1969} in the case of linear effects observed in Gaussian variables \citep{barnett2009}. 
Transfer Entropy based on conditional MI can be expressed as a difference of two MI terms:

\begin{equation} \label{eq_te_te}
T E(X, Y)_{\delta_{a}}=M I\left(X_{\delta_{a}},\left[Y_{-} Y_{0}\right]\right)-M I\left(X_{\delta_{a}}, Y_{-}\right)
\end{equation}

A critical problem is the concrete operationalisation of the target past $Y_-$.

\subsubsection*{One-dimensional embedding}

A simple form of operationalising the target past is to simply mirror the delay of the delay scanning procedure of the delayed MI \citep{besserve2010, lobier2014, park2015, ince2015, giordano2017, morillon2017}. 
When a given interaction delay is considered in diagonal TE, the delayed MI between $X$ and $Y$ is conditioned on a version of the target to which the same lag is applied as that which is applied to the source. 
We refer to this one-dimensional embedding as ``TE\textsubscript{1D}". 
With $Y_-$ thus corresponding to $Y_{\delta_a}$, we obtain:

\begin{equation} \label{eq_te_te1D}
T E_{\operatorname{1D}}(X, Y)_{\delta_{a}}=M I\left(X_{\delta_{a}},\left[Y_{\delta_{a}} Y_{0}\right]\right)-M I\left(X_{\delta_{a}}, Y_{\delta_{a}}\right)
\end{equation}

\subsubsection*{Multidimensional embedding approaches}

The one-dimensional method of operationalising the target past is simple to implement, but hard to motivate. 
When considering that the core goal of quantifying TE is to estimate to what degree the source variable contains information about the target that is not available from the target past itself, it becomes obvious that the assumption of capturing the target’s autoinformation with a single delay that varies as a function of the considered interaction delay is daunting \citep{wibral2013}. 
Multiple target past time points could jointly influence the target present, and there is no obvious reason why the target past should be varied as a function of the interaction delay considered. 
A source variable could have information about the target variable at a different delay than the delay at which the target past informs its own future in the same way as the source. 
A more sensible approach to describe the target past is thus to consider a multidimensional representation or ``embedding” \citep{takens1981} that is invariant to the interaction delay considered between source and target in the delay scanning procedure.
A classic idea to find such an embedding of the target past is parameterised by two parameters: 
A spacing parameter and a dimensionality parameter \citep{ragwitz2002}. 
This means that the assumption is that the information the target past contains about its present can be captured with a number of equally spaced delays relative to the present. 
A further development of embedding methods relaxes the assumption that the embedding delays must be uniformly spaced \citep{vlachos2010, faes2011}. 
It is conceivable that a lot of information about the target present can be found in a combination of many, densely clustered delays at very short time scales as well as only few unevenly spaced delays at longer time scales, or vice versa. 
To account for this, non-uniform embedding procedures adopt an iterative approach to generate a target past embedding. 
Concretely, a search space of candidate delays is defined by the maximal delay that is assumed to have an influence on the target present. 
In the first iteration, the delay within this search space is chosen that maximises the MI of the target variable at this delay relative to the target present. 
In subsequent iterations, a conditional MI between all remaining candidate delays and the target present conditioned on the already chosen delays is computed. 
If desired, it can be tested whether the found maximal conditional MI surpasses a defined noise threshold. 
In that case, the iterative procedure can be stopped once the found maximal conditional MI no longer exceeds the threshold. 
Alternatively, a pragmatic hyperparameter of maximal iterations can be defined, which avoids the computationally costly permutation testing. 
This procedure thus has two hyperparameters: the length of a search space as well as a noise threshold (in the form of a significance level alpha) or alternatively a fixed embedding dimensionality. 
In our study, we fixed these to a search space of 1 second and a fixed embedding dimensionality of 50 (simulations) or 25 (MEG data).
Conditional MI based TE estimators that employ such multidimensional target past operationalisations have been described to be  ``self prediction optimal” and  are thus referred to as TE\textsubscript{SPO} \citep{wibral2013}. 

\begin{equation} \label{eq_te_tespo}
T E_{S P O}(X, Y)_{\delta_{a}}=M I\left(X_{\delta_{a}},\left[Y_{e m b} Y_{0}\right]\right)-M I\left(X_{\delta_{a}}, Y_{e m b}\right)
\end{equation}

Here, $Y_{emb}$ denotes a multidimensional embedding of the target past.

\subsubsection*{Whitening approaches}
The suggestion we make here is to tackle the TE estimation problem with a whitening approach. 
The basic idea is to use a non-uniform multidimensional target past embedding as a model to derive a prediction about the target present state. 
Analogous to \citep{haugh1976}, this approach could then be used to subtract the prediction from the observed target time series and compute delayed MI on the residuals. 
This however ignores the uncertainty associated with each prediction. 
We thus instead compute a time series of sample wise conditional entropy of each sample in the observed target time series given its past embedding. 
We obtain the conditional entropy of each sample $x$ given its non-uniform embedding $x_{emb}$ by subtracting the marginal entropy of $x_{emb}$ from the joint entropy of $x$ and $x_{emb}$:

\begin{equation} \label{eq_te_ce}
h\left(x \mid x_{e m b}\right)=h\left(\left[x x_{e m b}\right]\right)-h\left(x_{e m b}\right)
\end{equation}

The resulting time series is thus effectively temporally decorrelated, or ``whitened”. 
We define DI\textsubscript{ce} between the source $X$ and the target $Y$ at an analysis delay $\delta_a$ as:

\begin{equation} \label{eq_te_DIce}
DI_{ce}(X, Y)_{\delta a}=M I\left(h\left(x \mid x_{e m b}\right)_{\delta_{a}}, h\left(y \mid y_{e m b}\right)\right)
\end{equation}

Any resulting quantifiable delayed MI between the whitened source and the target can thus no longer stem from information that the target carries about itself. 
To avoid asymmetrical temporal distortions caused by the temporal whitening, we apply the whitening to both source and target time series.

\subsection*{Partial information decomposition}
In Partial information decomposition \citep[PID,][]{williams2010}, systems of three or more variables are considered. 
In situations where there are two sources and one target variable, it is the goal to quantify the ``redundant” amount of information that the two sources share about the target as well as the ``synergistic” information about the target that is only available when considering the two sources jointly as well as the unique information about the target that is only available from one of the sources but not the other. 
From the lens of this formalism, conditional MI can be seen as the sum of two ``atoms” of PID: 
Unique information and synergy. 
TE based on conditional MI is thus the sum of not only unique information of the source signal about the target present but also synergy of the source signal and the target past about the target present.

PID has its origins in co-information \citep{mcgill1954}, which is computed as a triple set intersection.

\begin{equation} \label{eq_te_coi}
\operatorname{CoI}\left(X_{\delta_{a}}, Y_{e m b}, Y_{0}\right)=M I\left(X_{\delta_{a}}, Y_{0}\right)+M I\left(Y_{e m b}, Y_{0}\right)-M I\left(\left[X_{\delta_{a}} Y_{e m b}\right], Y_{0}\right)
\end{equation}

Here, negative values correspond to synergistic information, and positive values correspond to redundant information. 
From the lens of PID however, these are mere net sums of the ``pure” redundancy and synergy that PID aims to resolve. 
To do so, the essential first step is to define redundancy.
Here, we turned to an implementation based on ``common change in surprisal” \citep[$I_{ccs}$][]{ince2017}. 
This starts at the observation that local MI can be seen as the positive or negative change in surprisal of a given value when another is observed:

\begin{equation}
\operatorname{mi}(x, y)=\Delta_{y} h(x)=h(x)-h(x \mid y)
\end{equation}

We can then also consider equally positive or negative co-information at the local level:

\begin{equation}
\operatorname{coi}\left(x_{\delta_{a}}, y_{e m b}, y_{0}\right)=\operatorname{mi}\left(x_{\delta_{a}}, y_{0}\right)+m i\left(y_{e m b}, y_{0}\right)+m i\left(\left[x_{\delta_{a}} y_{e m b}\right], y_{0}\right)
\end{equation}

Each of such co-information terms can then either contribute to net synergy or net redundancy. 
The key idea in $I_{ccs}$ is then to consider only those positive (redundant) terms of local $coi(x_{\delta_a},y_{emb},y_0)$ which coincide with positive terms of $mi(x_{\delta_a},y_0)$, $mi(y_{emb},y_0)$ and $mi([x_{\delta_a} y_{emb}],y_0)$. Thus, only co-information terms are counted that represent a commonly shared change in surprisal. 
The resulting global redundancy can then be used to infer the other PID atoms according to a lattice structure \citep{williams2010}:

\begin{equation}
\begin{gathered}
\text { unique }\left(X_{\delta_{a}}\right)=M I\left(X_{\delta_{a}}, Y_{0}\right)-\text { redundancy } \\
\text { unique }\left(Y_{e m b}\right)=M I\left(Y_{e m b}, Y_{0}\right)-\text { redundancy } \\
\text { synergy }=M I\left(\left[X_{\delta_{a}} Y_{e m b}\right], Y_{0}\right)-\text { redundancy }-\text { unique }\left(X_{\delta_{a}}\right)-\text { unique }\left(Y_{e m b}\right)
\end{gathered}
\end{equation}

\subsection*{Noise thresholds}

To establish whether a given effect size exceeded a level that would be expected of data not containing the effect of interest, we considered the $95th$ percentile of distributions of effects obtained from $1000$ permutations.
For this, we performed circular shifts of the source variable by a random number of samples and then recomputed the estimators of interest. 
We constrained this random amount with a minimum and a maximum, such that the permuted data could not include instances where potential effects of the observed data would end up within the range of delays of interest. 
In cases where multiple comparisons were made, we corrected the noise threshold by means of the family wise error rate, that is, by considering the maximum across all conditions within a given permutation and subsequently applying the $95th$ percentile of the resulting distribution for all conditions.

\subsection*{Simulations}
The basic approach to simulating narrow-band time series was to firstly sample Gaussian white noise and subsequently apply band-pass filters to it ($3rd$ order butterworth, forward-only). 
Such signals $M$ were then circularly shifted by a ground truth interaction delay. 
Finally, source and target time series $X$ and $Y$ were generated by adding independent white noise of varying amplitude to $M$.
The following specific parameters were used for the individual simulations and analyses:

\begin{table}[!ht]
\centering
\sisetup{round-precision=3, add-integer-zero=false,table-figures-integer = 0,table-figures-decimal = 3, table-number-alignment = center, table-space-text-post = *}
\begin{tabular}{ c c c c c c c c}
\toprule
Figure & \makecell{Number \\of\\ samples} & \makecell{Noise \\amplitude \\source} & \makecell{Noise \\amplitude\\ target} & \makecell{Effect \\filter\\ $[$Hz$]$} & \makecell{Ground \\truth \\delay $[$s$]$} & \makecell{Analysis \\filter\\$[$Hz$]$} & \makecell{Lags $[$s$]$}\\
\midrule

{1} & $75000$ & $.5$ & $.25$ & $6\pm2$ & $-.12$ & $6\pm2$ & \makecell{$-.8$ - \\$+.32$ } \\

{2} & $75000$ & $.5$ & $.25$ & $6\pm2$ & $-.12$ & $6\pm2$ & \makecell{$-.8$ - \\$+.32$} \\

{3} & $40000$ & $.05$ & $.025$ & $6\pm2$ & \makecell{$-.08$ -\\ $-.4$} & $6\pm2$ & \makecell{$.66$ - \\$+.4$} \\

{4} & $40000$ & \makecell{$.001$ -\\ $10$} & \makecell{$.001$-\\$10$} &  $6\pm2$ & $-.12$ & $6\pm2$ & \makecell{$-.4$ - \\$+.2$} \\

{5} & $40000$ & \makecell{$0$ - \\$10$} & \makecell{$0$ -\\ $10$} & $6\pm2$ & $-.12$ & $6\pm2$ & \makecell{$-.4$ - \\$+.2$} \\

{6.1} & $40000$ & $.01$ & $.01$ & \makecell{$15 \pm1$ - \\$13.25$} & $-.12$ & \makecell{$15 \pm$1 -\\ $13.25$} & \makecell{$-.4$ - \\$+.2$} \\

{6.2} & $40000$ & $.01$ & $.01$ & $20 \pm5$ & $-.12$ & \makecell{$20 \pm1$ - \\$15$} & \makecell{$-.4$ - \\$+.2$} \\

{7.1} & $40000$ & $.01$ & $.01$ & \makecell{$3$ -\\ $50\pm2$} & $-.12$ & \makecell{$3$ -\\ $50\pm2$} & \makecell{$-.4$ - \\$+.2$}\\

{7.2} & $40000$ & $.01$ & $.01$ & \makecell{$3$ -\\ $47\pm2$} & $-.12$ & \makecell{$3$ -\\ $47\pm2$} & \makecell{$-.4$ - \\$+.2$}\\

{8} & $330000$ & $na$ & $na$ & $na$ & $na$ & \makecell{$1.5$ - \\$16 \pm1$} & \makecell{$-1$ - \\$.5$}\\

\bottomrule
\end{tabular}
\caption[Parameter settings in simulations.]{\textbf{Parameter settings in simulations.}
Figure 6.1 refers to panels A, B, D and E, Figure 6.2 refers to panels C and F. Figure 7.1 refers to panels A - D, Figure 7.2 refers to panels E - H. 
Effect and analysis filters are specified in terms of the passbands as defined by centre frequencies $\pm$ bandwidths.
$na$ denotes ``not available".
}
\label{table_te_parameters}
\end{table}

\subsection*{MEG data and analyses}

The MEG data in this study has been recorded and analysed before. 
For details concerning the participants, the experimental design, the recording procedures as well as the preprocessing, please refer to \citep{daube_simple_2019}.
In brief, $24$ participants had listened to an audiobook of $55$ minutes duration (in 6 blocks of equal duration) while their MEG (MAGNES 3600 WH, 4D Neuroimaging, $248$ magnetometers) had been recorded at a sampling rate of $1017.25$ Hz (first $10$ participants) or $2034.51$ Hz (last $14$ participants). 
We applied the same pre-processing steps as in the original study. 
This included interpolation of artifactual channels, replacement of squid jumps with DC patches, 4th order zero-phase high-pass filter of $.5$ Hz as well as independent component analysis for removal of eye and heart activity \citep{daube_simple_2019}. 
Here, we then downsampled the data to a sampling rate of $100$Hz.

To estimate activity from bilateral auditory cortices (ACs), we re-used linearly constrained minimum variance beamformer spatial filters \citep{vanveen1997} as in \citep{daube_simple_2019}. 
These had been optimised within a nested cross-validation to return responses that would be maximally correlated with linear predictions of the responses based on a combination of log-mel spectrograms and their temporal derivative. 
This correlation had been maximised with respect to position- and regularisation hyperparameters per participant, hemisphere and fold. 
Here, we averaged these hyperparameters across folds to then extract time series of activity from left and right ACs.

To define individual delta and theta frequency ranges, we considered spectra of delayed MI between AC activity and the speech envelope. 
To compute these spectra, we applied $3rd$ order butterworth forward filters to both AC activity and the envelope. 
These had centre frequencies increasing from $1.5$ to $15$ Hz in steps of $.25$ Hz, and cutoff frequencies were defined as bands of $\pm1$ Hz width around the centre frequencies. 
We computed delayed MI for a range of lags from $-1$ to $.5$s and used the maximum across lags within each frequency. 
We then searched for troughs in the resulting spectra and used the trough corresponding to the lowest frequency as the boundary between delta and theta. 
In $4$ out of $48$ cases, this returned boundaries between delta and theta that surpassed $6.5$ Hz, which we considered as unlikely a priori \citep{klimesch1999, wang2010}.
Therefore, in these cases we instead searched for peaks of the spectral gradient and used the peak corresponding to the lowest centre frequency. 
In practise, this corresponded to parts of the spectra where the initial decrease plateaued. 
To define an upper boundary of the theta component, we used the highest frequency at which the spectrum surpassed the noise threshold. 
This resulted in theta components with a broader bandwidth compared to the delta components. 
Since we had found the bandwidth of an effect to be correlated with the effect size recovered by DI\textsubscript{ce}, we also repeated analyses of delayed MI and DI\textsubscript{ce} with fixed bandwidths for delta and theta. 
For this, we defined the delta component to be below $3.5$ Hz and the theta component to be centered on a theta peak frequency, around which we defined bandpass filters of $3$ Hz width. 
The theta peak frequencies were defined as the maximum of the MI spectra within the theta frequency band as defined previously. 
In $4$ out of $48$ cases, this corresponded to the upper boundary of the delta band. In these cases, we defined the theta centre frequency to be $1.5$ Hz above the delta-theta boundary.

To extract time series corresponding to the passbands as defined above, we used $3rd$ order forward butterworth filters, which we applied to both the envelope and the AC activity. 
We then subjected the resulting time series to the computation of delayed MI and DI\textsubscript{ce}.
We statistically evaluated the nth observation of recovered log-transformed effect sizes and the recovered delays $r$ at the maxima across lags of each participant $p$, hemisphere $h$, frequency band $b$ and measure $m$ by fitting Bayesian linear distributional models as implemented in the brms package \citep{burkner2017}. 
These models can be summarised with the following formula:

\begin{equation}
\begin{aligned}
r_{[n]} &\sim N\left(\eta_{\mu[n]}, \exp \left(\eta_{\sigma[n]}\right)\right) \\
\eta_{\mu[n]} &\sim \beta_{\mu_{p[n]}}+\beta_{\mu_{p: h[n]}}+\beta_{\mu_{m: b[n]}} \\
\left(\beta_{\mu_{p[n]}}, \beta_{\mu_{p: h[n]^{\prime}}} \beta_{\mu_{m: b[n]}}\right) &\sim N(0, v) \\
\eta_{\sigma[n]} &\sim \beta_{\left.\sigma_{h[n]}\right]}+\beta_{\sigma_{m: b[n]}} \\
\left(\beta_{\left.\sigma_{h[n]}\right]}, \beta_{\sigma_{m: b[n]}}\right) &\sim N(0,10)
\end{aligned}
\end{equation}

\pagebreak
\beginsupplement
\section*{Supplementary Material}

\begin{figure}[!ht]
\centering
\includegraphics[width=\linewidth, trim={3cm 3.8cm 3.1cm 2.2cm},clip]{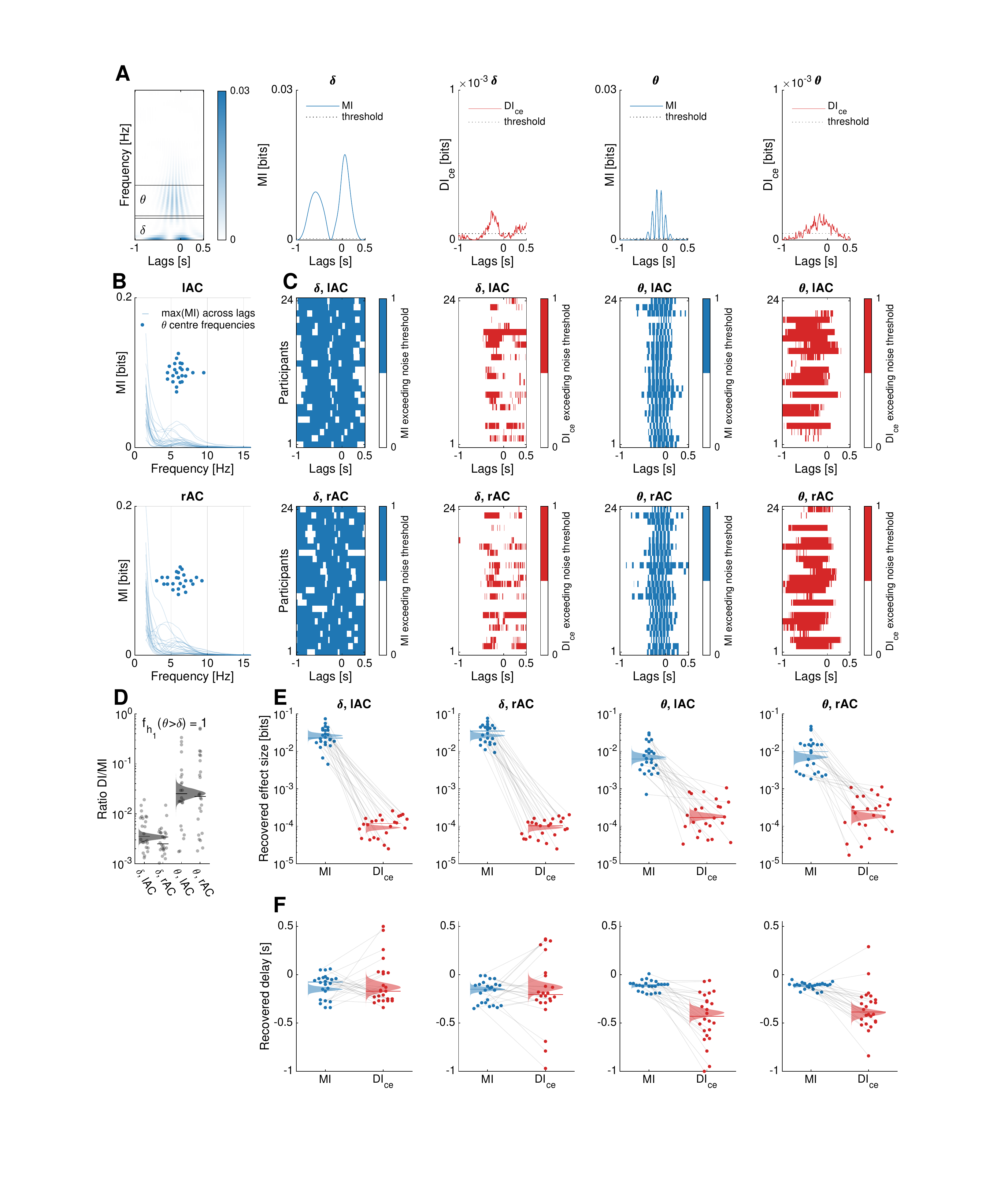}
\captionsetup{list=no}
\caption{Caption on following page.}
\end{figure}
\begin{figure}[!ht]
\captionsetup{labelformat=adja-page, list=yes}
\ContinuedFloat
\caption[Results obtained from source level MEG recordings with constrained frequencies.]{\textbf{Results obtained from source level MEG recordings with constrained frequencies (related to \hyperref[fig_te_meg]{figure \ref{fig_te_meg}}).}

Same as \hyperref[fig_te_meg]{figure \ref{fig_te_meg}}, but computed with constrained bandwidths of analysis filters.
Delta band was fixed by means of a low-pass filter common across all participants with a cutoff frequency of $3.5$Hz. 
Theta frequency was defined as a $3$Hz wide band centered on individual theta centre frequencies (see \hyperref[fig_te_sup_passbands]{figure \ref{fig_te_sup_passbands}}). 
The increased DI/MI ratio found with individualised band definitions (where often theta was defined as a wider band than delta) was robust to the constrained analysis bandwidths, making it unlikely that bandwidth alone can fully account for the difference in DI/MI ratio between delta and theta bands. 
Also see \hyperref[fig_te_sup_pp]{figure \ref{fig_te_sup_pp}}.
}
\label{fig_te_sup_megFixedFreq}
\end{figure}

\begin{figure}[!ht]
\centering
\includegraphics[width=\linewidth, trim={1.8cm .25cm .5cm .75cm},clip]{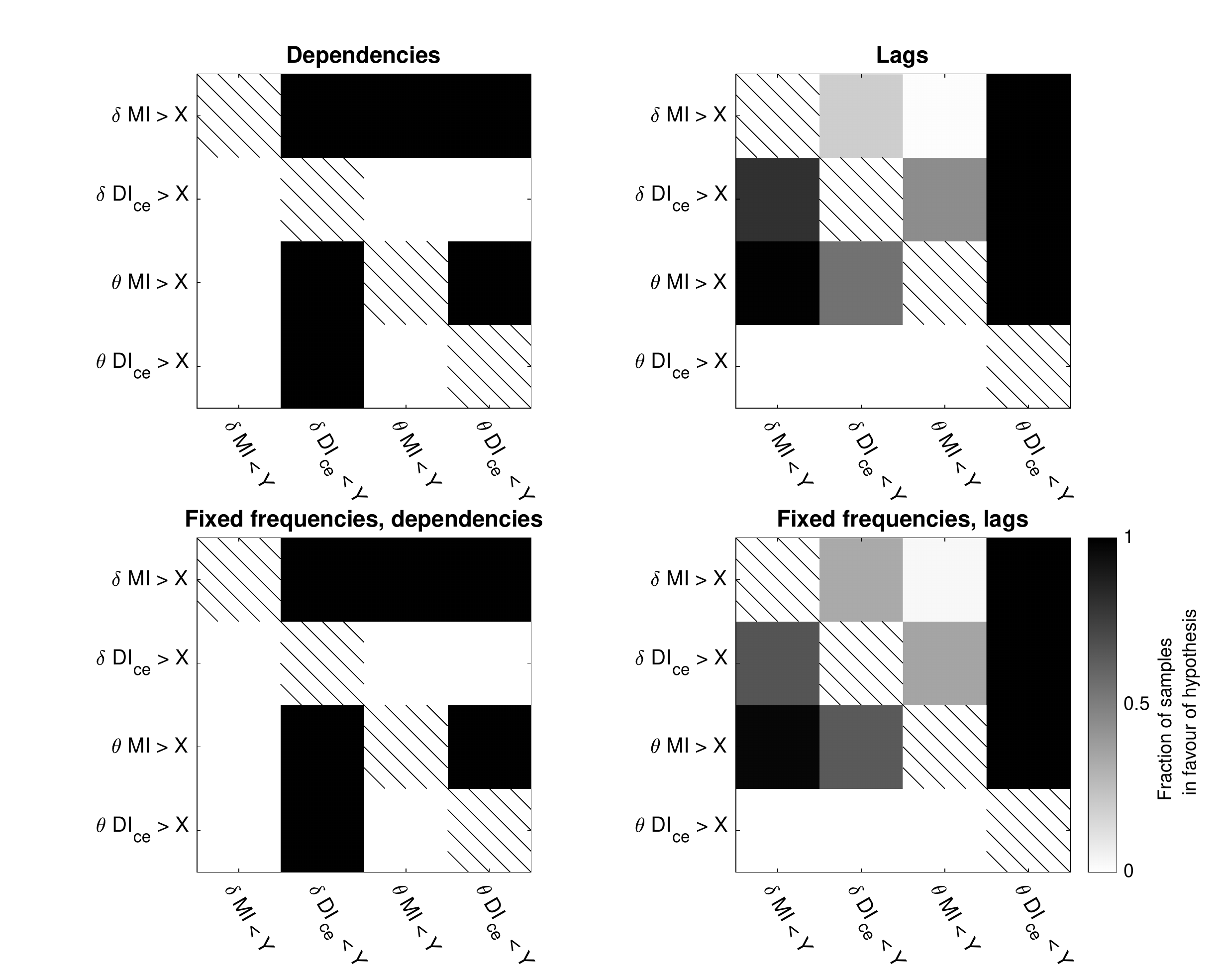}
\caption[Comparisons of posterior distributions of main effects of combinations of frequency bands and dependency measures from Bayesian linear modeling of the raw dependency and lag estimates.]{\textbf{Comparisons of posterior distributions of main effects of combinations of frequency bands and dependency measures from Bayesian linear modeling of the raw dependency and lag estimates (related to \hyperref[fig_te_meg]{figure \ref{fig_te_meg}}).}

In each cell in the matrices, the greyscale colour denotes the fraction of samples of the combination of frequency band and dependency measure referenced on the y-axis that is larger than the combination referenced on the x-axis (testing a hypothesis). 
Top row reports results corresponding to \hyperref[fig_te_meg]{figure \ref{fig_te_meg}}, bottom row reports results corresponding to \hyperref[fig_te_sup_megFixedFreq]{figure \ref{fig_te_sup_megFixedFreq}}.
}
\label{fig_te_sup_pp}
\end{figure}

\FloatBarrier
\pagebreak

\section*{Contributions}
\noindent C.D., J.G. and R.A.A.I. conceived of and designed the experiment.

\noindent C.D. collected and analyzed the data. 

\noindent C.D. and R.A.A.I. contributed analytic tools. 

\noindent C.D. wrote the manuscript.

\noindent C.D., J.G. and R.A.A.I. edited the manuscript. 

\section*{Acknowledgements}
R.A.A.I. was supported by the Wellcome Trust [214120/Z/18/Z].

\bibliographystyle{apacd}
\bibliography{bibliographie}

\begin{thebibliography}{}
\providecommand{\url}[1]{\texttt{#1}}
  \providecommand{\doi}[1]{doi: #1}
  \providecommand{\doi}{doi: \begingroup \urlstyle{rm}\Url}

\bibitem[\protect\astroncite{Ahissar et~al.}{2001}]{ahissar2001}
Ahissar, E., Nagarajan, S., Ahissar, M., Protopapas, A., Mahncke, H., \&
  Merzenich, M.~M. (2001).
\newblock Speech comprehension is correlated with temporal response patterns
  recorded from auditory cortex.
\newblock {\em Proceedings of the National Academy of Sciences},
  98(23):13367--13372.

\bibitem[\protect\astroncite{Barnett et~al.}{2009}]{barnett2009}
Barnett, L., Barrett, A.~B., \& Seth, A.~K. (2009).
\newblock Granger causality and transfer entropy are equivalent for Gaussian
  variables.
\newblock {\em Physical review letters}, 103(23):238701.

\bibitem[\protect\astroncite{Barnett \& Seth}{2011}]{barnett2011}
Barnett, L. \& Seth, A.~K. (2011).
\newblock Behaviour of Granger causality under filtering: theoretical
  invariance and practical application.
\newblock {\em Journal of neuroscience methods}, 201(2):404--419.

\bibitem[\protect\astroncite{Barrett}{2015}]{barrett2015}
Barrett, A.~B. (2015).
\newblock Exploration of synergistic and redundant information sharing in
  static and dynamical Gaussian systems.
\newblock {\em Physical Review E}, 91(5):052802.

\bibitem[\protect\astroncite{Bassett \& Bullmore}{2006}]{bassett2006}
Bassett, D.~S. \& Bullmore, E. (2006).
\newblock Small-world brain networks.
\newblock {\em The neuroscientist}, 12(6):512--523.

\bibitem[\protect\astroncite{Bastos \& Schoffelen}{2016}]{bastos2016}
Bastos, A.~M. \& Schoffelen, J.-M. (2016).
\newblock A tutorial review of functional connectivity analysis methods and
  their interpretational pitfalls.
\newblock {\em Frontiers in systems neuroscience}, 9:175.

\bibitem[\protect\astroncite{Besserve et~al.}{2010}]{besserve2010}
Besserve, M., Sch{\"o}lkopf, B., Logothetis, N.~K., \& Panzeri, S. (2010).
\newblock Causal relationships between frequency bands of extracellular signals
  in visual cortex revealed by an information theoretic analysis.
\newblock {\em Journal of computational neuroscience}, 29(3):547--566.

\bibitem[\protect\astroncite{Brodbeck et~al.}{2018}]{brodbeck2018b}
Brodbeck, C., Hong, L.~E., \& Simon, J.~Z. (2018).
\newblock Rapid Transformation from Auditory to Linguistic Representations of
  Continuous Speech.
\newblock {\em Current Biology}, 28:3976--3983.

\bibitem[\protect\astroncite{Brookes et~al.}{2011}]{brookes2011}
Brookes, M.~J., Woolrich, M., Luckhoo, H., Price, D., Hale, J.~R., Stephenson,
  M.~C., Barnes, G.~R., Smith, S.~M., \& Morris, P.~G. (2011).
\newblock Investigating the electrophysiological basis of resting state
  networks using magnetoencephalography.
\newblock {\em Proceedings of the National Academy of Sciences},
  108(40):16783--16788.

\bibitem[\protect\astroncite{B{\"u}rkner}{2017}]{burkner2017}
B{\"u}rkner, P.~C. (2017).
\newblock brms: An {R} Package for Bayesian Multilevel Models Using Stan.
\newblock {\em Journal of Statistical Software}, 80(1).

\bibitem[\protect\astroncite{Caucheteux et~al.}{2021}]{caucheteux2021b}
Caucheteux, C., Gramfort, A., \& King, J.-R. (2021).
\newblock Long-range and hierarchical language predictions in brains and
  algorithms.
\newblock {\em arXiv preprint arXiv:2111.14232}.

\bibitem[\protect\astroncite{Chung et~al.}{2020}]{chung2020}
Chung, Y.-A., Tang, H., \& Glass, J. (2020).
\newblock Vector-quantized autoregressive predictive coding.
\newblock {\em arXiv preprint arXiv:2005.08392}.

\bibitem[\protect\astroncite{Cliff et~al.}{2021}]{cliff2021}
Cliff, O.~M., Novelli, L., Fulcher, B.~D., Shine, J.~M., \& Lizier, J.~T.
  (2021).
\newblock Assessing the significance of directed and multivariate measures of
  linear dependence between time series.
\newblock {\em Physical Review Research}, 3(1):013145.

\bibitem[\protect\astroncite{Crosse et~al.}{2016}]{crosse2016}
Crosse, M.~J., DiLiberto, G.~M., Bednar, A., \& Lalor, E.~C. (2016).
\newblock The Multivariate Temporal Response Function (m{TRF}) Toolbox: A
  MATLAB Toolbox for Relating Neural Signals to Continuous Stimuli.
\newblock {\em frontiers in Human Neuroscience}, 10(604).

\bibitem[\protect\astroncite{Daube et~al.}{2019}]{daube_simple_2019}
Daube, C., Ince, R. A.~A., \& Gross, J. (2019).
\newblock Simple {Acoustic} {Features} {Can} {Explain} {Phoneme}-{Based}
  {Predictions} of {Cortical} {Responses} to {Speech}.
\newblock {\em Current Biology}, 29(12):1924--1937.e9.
\newblock \doi{10.1016/j.cub.2019.04.067}.
\newblock URL
  \url{https://www.sciencedirect.com/science/article/pii/S0960982219304968}.

\bibitem[\protect\astroncite{{de-Wit} et~al.}{2016}]{dewit2016}
{de-Wit}, L., Alexander, D., Ekroll, V., \& Wagemans, J. (2016).
\newblock Is neuroimaging measuring information in the brain?
\newblock {\em Psychonomic Bulletin \& Reviews}, 23:1415---1428.

\bibitem[\protect\astroncite{Di~Liberto et~al.}{2015}]{diliberto2015}
Di~Liberto, G.~M., O'Sullivan, J.~A., \& Lalor, E.~C. (2015).
\newblock Low-Frequency Cortical Entrainment to Speech Reflects Phoneme-Level
  Processing.
\newblock {\em Current Biology}, 25:2457--2465.

\bibitem[\protect\astroncite{Ding et~al.}{2016}]{ding2016}
Ding, N., Melloni, L., Zhang, H., Tian, X., \& Poeppel, D. (2016).
\newblock Cortical tracking of hierarchical linguistic structures in connected
  speech.
\newblock {\em Nature Neuroscience}, 19(1):158--164.

\bibitem[\protect\astroncite{Ding \& Simon}{2012}]{ding2012}
Ding, N. \& Simon, J.~Z. (2012).
\newblock Emergence of neural encoding of auditory objects while listening to
  competing speakers.
\newblock {\em Proceedings of the National Academy of Sciences of the United
  States of America}, 109(29):11854--11859.

\bibitem[\protect\astroncite{Ding \& Simon}{2014}]{ding2014}
Ding, N. \& Simon, J.~Z. (2014).
\newblock Cortical entrainment to continuous speech: functional roles and
  interpretations.
\newblock {\em Frontiers in human neuroscience}, 8:311.

\bibitem[\protect\astroncite{Donhauser \& Baillet}{2020}]{donhauser2020}
Donhauser, P.~W. \& Baillet, S. (2020).
\newblock Two distinct neural timescales for predictive speech processing.
\newblock {\em Neuron}, 105(2):385--393.

\bibitem[\protect\astroncite{Donoghue et~al.}{2020}]{donoghue2020}
Donoghue, T., Haller, M., Peterson, E.~J., Varma, P., Sebastian, P., Gao, R.,
  Noto, T., Lara, A.~H., Wallis, J.~D., Knight, R.~T., et~al. (2020).
\newblock Parameterizing neural power spectra into periodic and aperiodic
  components.
\newblock {\em Nature neuroscience}, 23(12):1655--1665.

\bibitem[\protect\astroncite{Etard \& Reichenbach}{2019}]{etard2019}
Etard, O. \& Reichenbach, T. (2019).
\newblock Neural speech tracking in the theta and in the delta frequency band
  differentially encode clarity and comprehension of speech in noise.
\newblock {\em Journal of Neuroscience}, 39(29):5750--5759.

\bibitem[\protect\astroncite{Faes et~al.}{2011}]{faes2011}
Faes, L., Nollo, G., \& Porta, A. (2011).
\newblock Information-based detection of nonlinear Granger causality in
  multivariate processes via a nonuniform embedding technique.
\newblock {\em Physical Review E}, 83(5):051112.

\bibitem[\protect\astroncite{Faes et~al.}{2017}]{faes2017}
Faes, L., Nollo, G., Stramaglia, S., \& Marinazzo, D. (2017).
\newblock Multiscale granger causality.
\newblock {\em Physical Review E}, 96(4):042150.

\bibitem[\protect\astroncite{Florin et~al.}{2010}]{florin2010}
Florin, E., Gross, J., Pfeifer, J., Fink, G.~R., \& Timmermann, L. (2010).
\newblock The effect of filtering on Granger causality based multivariate
  causality measures.
\newblock {\em Neuroimage}, 50(2):577--588.

\bibitem[\protect\astroncite{Fries}{2015}]{fries2015}
Fries, P. (2015).
\newblock Rhythms for cognition: communication through coherence.
\newblock {\em Neuron}, 88(1):220--235.

\bibitem[\protect\astroncite{Friston}{2005}]{friston2005}
Friston, K. (2005).
\newblock A theory of cortical responses.
\newblock {\em Philosophical transactions of the Royal Society B: Biological
  sciences}, 360(1456):815--836.

\bibitem[\protect\astroncite{Gerster et~al.}{2021}]{gerster2021}
Gerster, M., Waterstraat, G., Litvak, V., Lehnertz, K., Schnitzler, A., Florin,
  E., Curio, G., \& Nikulin, V. (2021).
\newblock Separating neural oscillations from aperiodic 1/f activity:
  challenges and recommendations.
\newblock {\em bioRxiv}.

\bibitem[\protect\astroncite{Giordano et~al.}{2016}]{giordano2017}
Giordano, B.~L., Ince, R. A.~A., Gross, J., Schyns, P.~G., Panzeri, S., \&
  Kayser, C. (2016).
\newblock Contributions of local speech encoding and functional connectivity to
  audio-visual speech perception.
\newblock {\em eLife}, 6:e24763(DOI: 10.7554/eLife.24763).

\bibitem[\protect\astroncite{Giraud \& Poeppel}{2012}]{giraud2012}
Giraud, A.~L. \& Poeppel, D. (2012).
\newblock Cortical oscillations and speech processing: emerging computational
  principles and operations.
\newblock {\em Nature Neuroscience}, 15:511--517.

\bibitem[\protect\astroncite{Granger}{1969}]{granger1969}
Granger, C.~W. (1969).
\newblock Investigating causal relations by econometric models and
  cross-spectral methods.
\newblock {\em Econometrica: journal of the Econometric Society}, pages
  424--438.

\bibitem[\protect\astroncite{Gross et~al.}{2013}]{gross2013}
Gross, J., Hoogenboom, N., Thut, G., Schyns, P.~G., Panzeri, S., Belin, P., \&
  Garrod, S. (2013).
\newblock Speech Rhythms and Multiplexed Oscillatory Sensory Coding in the
  Human Brain.
\newblock {\em PLoS Biology}, 11(12).

\bibitem[\protect\astroncite{Gross et~al.}{2021}]{gross2021}
Gross, J., Kluger, D.~S., Abbasi, O., Chalas, N., Steingr{\"a}ber, N., Daube,
  C., \& Schoffelen, J.-M. (2021).
\newblock Comparison of undirected frequency-domain connectivity measures for
  cerebro-peripheral analysis.
\newblock {\em NeuroImage}, 245:118660.

\bibitem[\protect\astroncite{Haugh}{1976}]{haugh1976}
Haugh, L.~D. (1976).
\newblock Checking the independence of two covariance-stationary time series: a
  univariate residual cross-correlation approach.
\newblock {\em Journal of the American Statistical Association},
  71(354):378--385.

\bibitem[\protect\astroncite{Heilbron et~al.}{2021}]{heilbron2021}
Heilbron, M., Armeni, K., Schoffelen, J.-M., Hagoort, P., \& de~Lange, F.~P.
  (2021).
\newblock A hierarchy of linguistic predictions during natural language
  comprehension.
\newblock {\em bioRxiv}, pages 2020--12.

\bibitem[\protect\astroncite{Henry \& Obleser}{2012}]{henry2012}
Henry, M.~J. \& Obleser, J. (2012).
\newblock Frequency modulation entrains slow neural oscillations and optimizes
  human listening behavior.
\newblock {\em Proceedings of the National Academy of Sciences},
  109(49):20095--20100.

\bibitem[\protect\astroncite{Hertrich et~al.}{2012}]{hertrich2012}
Hertrich, I., Dietrich, S., Trouvain, J., Moos, A., \& Ackermann, H. (2012).
\newblock Magnetic brain activity phase-locked to the envelope, the syllable
  onsets, and the fundamental frequency of a perceived speech signal.
\newblock {\em Psychophysiology}, 49:322--334.

\bibitem[\protect\astroncite{Hyafil et~al.}{2015}]{hyafil2015}
Hyafil, A., Fontolan, L., Kabdebon, C., Boris, G., \& Giraud, A.~L. (2015).
\newblock Speech encoding by coupled cortical theta and gamma oscillations.
\newblock {\em eLife}, 4(e06213).

\bibitem[\protect\astroncite{Ince}{2017}]{ince2017}
Ince, R. (2017).
\newblock Measuring {Multivariate} {Redundant} {Information} with {Pointwise}
  {Common} {Change} in {Surprisal}.
\newblock {\em Entropy}, 19(7):318.
\newblock \doi{10.3390/e19070318}.
\newblock URL \url{http://www.mdpi.com/1099-4300/19/7/318}.

\bibitem[\protect\astroncite{Ince et~al.}{2015}]{ince2015}
Ince, R.~A., Van~Rijsbergen, N.~J., Thut, G., Rousselet, G.~A., Gross, J.,
  Panzeri, S., \& Schyns, P.~G. (2015).
\newblock Tracing the flow of perceptual features in an algorithmic brain
  network.
\newblock {\em Scientific reports}, 5(1):1--17.

\bibitem[\protect\astroncite{Ince et~al.}{2017}]{ince2017b}
Ince, R. A.~A., Giordano, B.~L., Kayser, C., Rousselet, G.~A., Gross, J., \&
  Schyns, P.~G. (2017).
\newblock A Statistical Framework for Neuroimaging Data Analysis Based on
  Mutual Information Estimated via a Gaussian Copula.
\newblock {\em Human Brain Mapping}, 38(3):1541--1573.

\bibitem[\protect\astroncite{Ince et~al.}{2021}]{ince_bayesian_2021}
Ince, R. A.~A., Kay, J.~W., \& Schyns, P.~G. (2021).
\newblock Bayesian inference of population prevalence.
\newblock {\em eLife}, 10(e62461).
\newblock \doi{10.7554/eLife.62461}.
\newblock URL \url{https://elifesciences.org/articles/62461}.

\bibitem[\protect\astroncite{Jain et~al.}{2021}]{jain2021}
Jain, S., Mahto, S., Turek, J.~S., Vo, V.~A., LeBel, A., \& Huth, A.~G. (2021).
\newblock Interpretable multi-timescale models for predicting fMRI responses to
  continuous natural speech.
\newblock {\em bioRxiv}, pages 2020--10.

\bibitem[\protect\astroncite{James et~al.}{2016}]{james2016}
James, R.~G., Barnett, N., \& Crutchfield, J.~P. (2016).
\newblock Information flows? A critique of transfer entropies.
\newblock {\em Physical review letters}, 116(23):238701.

\bibitem[\protect\astroncite{Jones \& Boltz}{1989}]{jones1989}
Jones, M.~R. \& Boltz, M. (1989).
\newblock Dynamic attending and responses to time.
\newblock {\em Psychological review}, 96(3):459.

\bibitem[\protect\astroncite{Kayser et~al.}{2015}]{kayser2015}
Kayser, C., Wilson, C., Safaai, H., Sakata, S., \& Panzeri, S. (2015).
\newblock Rhythmic auditory cortex activity at multiple timescales shapes
  stimulus--response gain and background firing.
\newblock {\em Journal of Neuroscience}, 35(20):7750--7762.

\bibitem[\protect\astroncite{Klimesch}{1999}]{klimesch1999}
Klimesch, W. (1999).
\newblock EEG alpha and theta oscillations reflect cognitive and memory
  performance: a review and analysis.
\newblock {\em Brain research reviews}, 29(2-3):169--195.

\bibitem[\protect\astroncite{Koskinen et~al.}{2020}]{koskinen2020}
Koskinen, M., Kurimo, M., Gross, J., Hyv{\"a}rinen, A., \& Hari, R. (2020).
\newblock Brain activity reflects the predictability of word sequences in
  listened continuous speech.
\newblock {\em NeuroImage}, 219:116936.

\bibitem[\protect\astroncite{Kulasingham et~al.}{2020}]{kulasingham2020}
Kulasingham, J.~P., Brodbeck, C., Presacco, A., Kuchinsky, S.~E., Anderson, S.,
  \& Simon, J.~Z. (2020).
\newblock High gamma cortical processing of continuous speech in younger and
  older listeners.
\newblock {\em Neuroimage}, 222:117291.

\bibitem[\protect\astroncite{Lakatos et~al.}{2019}]{lakatos2019}
Lakatos, P., Gross, J., \& Thut, G. (2019).
\newblock A new unifying account of the roles of neuronal entrainment.
\newblock {\em Current Biology}, 29(18):R890--R905.

\bibitem[\protect\astroncite{Lakatos et~al.}{2008}]{lakatos2008}
Lakatos, P., Karmos, G., Mehta, A.~D., Ulbert, I., \& Schroeder, C.~E. (2008).
\newblock Entrainment of neuronal oscillations as a mechanism of attentional
  selection.
\newblock {\em science}, 320(5872):110--113.

\bibitem[\protect\astroncite{Lakhotia et~al.}{2021}]{lakhotia2021}
Lakhotia, K., Kharitonov, E., Hsu, W.-N., Adi, Y., Polyak, A., Bolte, B.,
  Nguyen, T.-A., Copet, J., Baevski, A., Mohamed, A., et~al. (2021).
\newblock Generative spoken language modeling from raw audio.
\newblock {\em arXiv preprint arXiv:2102.01192}.

\bibitem[\protect\astroncite{Lalor et~al.}{2009}]{lalor2009}
Lalor, E.~C., Power, A.~J., Reilly, R.~B., \& Foxe, J.~J. (2009).
\newblock Resolving precise temporal processing properties of the auditory
  system using continuous stimuli.
\newblock {\em Journal of neurophysiology}, 102(1):349--359.

\bibitem[\protect\astroncite{Lobier et~al.}{2014}]{lobier2014}
Lobier, M., Siebenh{\"u}hner, F., Palva, S., \& Palva, J.~M. (2014).
\newblock Phase transfer entropy: a novel phase-based measure for directed
  connectivity in networks coupled by oscillatory interactions.
\newblock {\em Neuroimage}, 85:853--872.

\bibitem[\protect\astroncite{Makeig et~al.}{2002}]{makeig2002}
Makeig, S., Westerfield, M., Jung, T.-P., Enghoff, S., Townsend, J.,
  Courchesne, E., \& Sejnowski, T.~J. (2002).
\newblock Dynamic brain sources of visual evoked responses.
\newblock {\em Science}, 295(5555):690--694.

\bibitem[\protect\astroncite{McGill}{1954}]{mcgill1954}
McGill, W.~J. (1954).
\newblock Multivariate information transmission.
\newblock {\em Psychometrika}, 19:97--116.

\bibitem[\protect\astroncite{Mehler \& Kording}{2018}]{mehler2018}
Mehler, D. M.~A. \& Kording, K.~P. (2018).
\newblock The lure of causal statements: Rampant mis-inference of causality in
  estimated connectivity.
\newblock {\em arXiv e-prints}, pages arXiv--1812.

\bibitem[\protect\astroncite{Mell et~al.}{2021}]{mell2021}
Mell, M.~M., St-Yves, G., \& Naselaris, T. (2021).
\newblock Voxel-to-voxel predictive models reveal unexpected structure in
  unexplained variance.
\newblock {\em NeuroImage}, page 118266.

\bibitem[\protect\astroncite{Michalareas et~al.}{2016}]{michalareas2016}
Michalareas, G., Vezoli, J., Van~Pelt, S., Schoffelen, J.-M., Kennedy, H., \&
  Fries, P. (2016).
\newblock Alpha-beta and gamma rhythms subserve feedback and feedforward
  influences among human visual cortical areas.
\newblock {\em Neuron}, 89(2):384--397.

\bibitem[\protect\astroncite{M{\l}ynarski \&
  Hermundstad}{2018}]{mlynarski2018b}
M{\l}ynarski, W.~F. \& Hermundstad, A.~M. (2018).
\newblock Adaptive coding for dynamic sensory inference.
\newblock {\em Elife}, 7:e32055.

\bibitem[\protect\astroncite{Morillon \& Baillet}{2017}]{morillon2017}
Morillon, B. \& Baillet, S. (2017).
\newblock Motor origin of temporal predictions in auditory attention.
\newblock {\em Proceedings of the National Academy of Sciences},
  114(42):E8913--E8921.

\bibitem[\protect\astroncite{Naselaris et~al.}{2011}]{naselaris2011}
Naselaris, T., Kay, K.~N., Nishimoto, S., \& Gallant, J.~L. (2011).
\newblock Encoding and decoding in {fMRI}.
\newblock {\em NeuroImage}, 56(2):400--410.
\newblock \doi{10.1016/j.neuroimage.2010.07.073}.
\newblock URL
  \url{http://www.sciencedirect.com/science/article/pii/S1053811910010657}.

\bibitem[\protect\astroncite{Obleser \& Kayser}{2019}]{obleser2019}
Obleser, J. \& Kayser, C. (2019).
\newblock Neural entrainment and attentional selection in the listening brain.
\newblock {\em Trends in cognitive sciences}, 23(11):913--926.

\bibitem[\protect\astroncite{O'Sullivan et~al.}{2015}]{osullivan2015}
O'Sullivan, J.~A., Power, A.~J., Mesgarani, N., Rajaram, S., Foxe, J.~J.,
  Shinn-Cunningham, B.~G., Slaney, M., Shamma, S.~A., \& Lalor, E.~C. (2015).
\newblock Attentional selection in a cocktail party environment can be decoded
  from single-trial EEG.
\newblock {\em Cerebral cortex}, 25(7):1697--1706.

\bibitem[\protect\astroncite{Park et~al.}{2015}]{park2015}
Park, H., Ince, R.~A., Schyns, P.~G., Thut, G., \& Gross, J. (2015).
\newblock Frontal top-down signals increase coupling of auditory low-frequency
  oscillations to continuous speech in human listeners.
\newblock {\em Current Biology}, 25(12):1649--1653.

\bibitem[\protect\astroncite{Pinzuti et~al.}{2020}]{pinzuti2020}
Pinzuti, E., Wollstadt, P., Gutknecht, A., T{\"u}scher, O., \& Wibral, M.
  (2020).
\newblock Measuring spectrally-resolved information transfer.
\newblock {\em PLoS Computational Biology}, 16(12):e1008526.

\bibitem[\protect\astroncite{Ragwitz \& Kantz}{2002}]{ragwitz2002}
Ragwitz, M. \& Kantz, H. (2002).
\newblock Markov models from data by simple nonlinear time series predictors in
  delay embedding spaces.
\newblock {\em Physical Review E}, 65(5):056201.

\bibitem[\protect\astroncite{Rao \& Ballard}{1999}]{rao1999}
Rao, R.~P. \& Ballard, D.~H. (1999).
\newblock Predictive coding in the visual cortex: a functional interpretation
  of some extra-classical receptive-field effects.
\newblock {\em Nature neuroscience}, 2(1):79--87.

\bibitem[\protect\astroncite{Ritter et~al.}{1968}]{ritter1968}
Ritter, W., Vaughan~Jr, H.~G., \& Costa, L.~D. (1968).
\newblock Orienting and habituation to auditory stimuli: a study of short terms
  changes in average evoked responses.
\newblock {\em Electroencephalography and clinical Neurophysiology},
  25(6):550--556.

\bibitem[\protect\astroncite{Sayers et~al.}{1974}]{sayers1974}
Sayers, B.~M., Beagley, H.~A., \& Henshall, W.~R. (1974).
\newblock The mechanism of auditory evoked EEG responses.
\newblock {\em Nature}, 247(5441):481--483.

\bibitem[\protect\astroncite{Schmitt et~al.}{2021}]{schmitt2021}
Schmitt, L.-M., Erb, J., Tune, S., Rysop, A., Hartwigsen, G., \& Obleser, J.
  (2021).
\newblock Predicting speech from a cortical hierarchy of event-based time
  scales.
\newblock {\em Science Advances}, 7(49).

\bibitem[\protect\astroncite{Schnitzler \& Gross}{2005}]{schnitzler2005}
Schnitzler, A. \& Gross, J. (2005).
\newblock Normal and pathological oscillatory communication in the brain.
\newblock {\em Nature reviews neuroscience}, 6(4):285--296.

\bibitem[\protect\astroncite{Schoffelen et~al.}{2017}]{schoffelen2017}
Schoffelen, J.-M., Hult{\'e}n, A., Lam, N., Marquand, A.~F., Udd{\'e}n, J., \&
  Hagoort, P. (2017).
\newblock Frequency-specific directed interactions in the human brain network
  for language.
\newblock {\em Proceedings of the National Academy of Sciences},
  114(30):8083--8088.

\bibitem[\protect\astroncite{Schreiber}{2000}]{schreiber2000}
Schreiber, T. (2000).
\newblock Measuring information transfer.
\newblock {\em Physical review letters}, 85(2):461.

\bibitem[\protect\astroncite{Schroeder \& Lakatos}{2009}]{schroeder2009}
Schroeder, C.~E. \& Lakatos, P. (2009).
\newblock Low-frequency neuronal oscillations as instruments of sensory
  selection.
\newblock {\em Trends in neurosciences}, 32(1):9--18.

\bibitem[\protect\astroncite{Takens}{1981}]{takens1981}
Takens, F. (1981).
\newblock Detecting strange attractors in turbulence.
\newblock In {\em Dynamical systems and turbulence, Warwick 1980}, pages
  366--381. Springer.

\bibitem[\protect\astroncite{Van~Veen et~al.}{1997}]{vanveen1997}
Van~Veen, B.~D., van Drongelen, W., Yuchtman, M., \& Suzuki, A. (1997).
\newblock Localization of Brain Electrical Activity via Linearly Constrained
  Minimum Variance Spatial Filtering.
\newblock {\em IEEE Transactions on Biomedical Engineering}, 44(9):867--880.

\bibitem[\protect\astroncite{Vlachos \& Kugiumtzis}{2010}]{vlachos2010}
Vlachos, I. \& Kugiumtzis, D. (2010).
\newblock Nonuniform state-space reconstruction and coupling detection.
\newblock {\em Physical Review E}, 82(1):016207.

\bibitem[\protect\astroncite{Wang}{2010}]{wang2010}
Wang, X.-J. (2010).
\newblock Neurophysiological and computational principles of cortical rhythms
  in cognition.
\newblock {\em Physiological reviews}, 90(3):1195--1268.

\bibitem[\protect\astroncite{Wibral et~al.}{2013}]{wibral2013}
Wibral, M., Pampu, N., Priesemann, V., Siebenh{\"u}hner, F., Seiwert, H.,
  Lindner, M., Lizier, J.~T., \& Vicente, R. (2013).
\newblock Measuring information-transfer delays.
\newblock {\em PloS one}, 8(2):e55809.

\bibitem[\protect\astroncite{Williams \& Beer}{2010}]{williams2010}
Williams, P.~L. \& Beer, R.~D. (2010).
\newblock Nonnegative Decomposition of Multivariate Information.
\newblock {\em arXiv}, 1004.2515.

\bibitem[\protect\astroncite{Wollstadt et~al.}{2017}]{wollstadt2017}
Wollstadt, P., Sellers, K.~K., Rudelt, L., Priesemann, V., Hutt, A.,
  Fr{\"o}hlich, F., \& Wibral, M. (2017).
\newblock Breakdown of local information processing may underlie isoflurane
  anesthesia effects.
\newblock {\em PLoS computational biology}, 13(6):e1005511.

\bibitem[\protect\astroncite{W{\"o}stmann et~al.}{2017}]{wostmann2017}
W{\"o}stmann, M., Fiedler, L., \& Obleser, J. (2017).
\newblock Tracking the signal, cracking the code: Speech and speech
  comprehension in non-invasive human electrophysiology.
\newblock {\em Language, Cognition and Neuroscience}, 32(7):855--869.

\bibitem[\protect\astroncite{Zan et~al.}{2020}]{zan2020}
Zan, P., Presacco, A., Anderson, S., \& Simon, J.~Z. (2020).
\newblock Exaggerated cortical representation of speech in older listeners:
  mutual information analysis.
\newblock {\em Journal of Neurophysiology}, 124(4):1152--1164.

\end{thebibliography}
\end{document}